\documentclass[epj]{svjour} 

\usepackage[utf8]{inputenc}
\usepackage{graphicx}
\usepackage{amsmath}
\usepackage{amsfonts}
\usepackage{ bbold }
\usepackage{booktabs}

\newcommand{\beq}{\begin{equation}}
\newcommand{\eeq}{\end{equation}}
\newcommand{\beqa}{\begin{eqnarray}}
\newcommand{\eeqa}{\end{eqnarray}}
\def\lsim{\raise0.3ex\hbox{$<$\kern-0.75em\raise-1.1ex\hbox{$\sim$}}}
\def\gsim{\raise0.3ex\hbox{$>$\kern-0.75em\raise-1.1ex\hbox{$\sim$}}}

\usepackage{color} % colorful text
\def\blr#1{\left(#1\right)}
\def\slr#1{\left[#1\right]}
\newcommand{\be}{\begin{equation}}
\newcommand{\ee}{\end{equation}}
\newcommand{\bea}{\begin{eqnarray}}
\newcommand{\eea}{\end{eqnarray}}

\def\Fig#1{Fig.~\ref{#1}}
\def\mc#1{\mathcal{#1}}
\newcommand{\mrm}[1]{\mathrm{#1}}
\newcommand{\lone}{\lambda_1}
\newcommand{\ltwo}{\lambda_2}
\def\Nc{N_\mrm{c}}
\def\pd{\partial}
\def\Phib{\bar{\Phi}}

\def\s{\mrm{s}}
\def\sbar{\bar{\sigma}}

\newcommand{\Tr}{\mathrm{Tr}}
\DeclareMathOperator{\e}{e}

\begin{document}
\title{Isentropic evolution of the matter in heavy-ion collisions and the search for the critical endpoint}
\author{Mario Motta\inst{1,2} \and Rainer Stiele\inst{2,3,4} \and Wanda Maria Alberico\inst{1,2} \and Andrea Beraudo\inst{2}}

\institute{Dipartimento di Fisica, Universit\`a degli Studi di Torino, via Pietro Giuria 1, I-10125 - Torino, Italy \and INFN, Sezione di Torino, via Pietro Giuria 1, I-10125 Torino, Italy \and Univ~Lyon, ENS de Lyon, D\'epartement de Physique, F-69342, Lyon, France \and Univ~Lyon, Univ Claude Bernard Lyon 1, CNRS/IN2P3, IP2I Lyon, F-69622, Villeurbanne, France}
\date{}

\abstract{We study the isentropic evolution of the matter produced in relativistic heavy-ion collisions for various values of the entropy-per-baryon ratio of interest for the ongoing and future experimental searches for the critical endpoint (CEP) in the QCD phase diagram: these includes the current Beam-Energy-Scan (BEC) program at RHIC and the fixed-target collisions foreseen for the near future at various facilities. We describe the hot-dense matter through two different effective Lagrangians: the PNJL (Polyakov-Nambu-Jona-Lasinio) and the PQM (Polyakov-quark-meson) models. We focus on quantities expected to have a direct experimental relevance: the speed of sound, responsible for the collective acceleration of the fireball, and the generalized susceptibilities, connected to the cumulants of the distributions of conserved charges. In principle they should affect the momentum spectra and the event-by-event fluctuations of the yields of identified particles. Taking realistic values for the initial temperature and the entropy-per-baryon ratio we study the temporal evolution of the above quantities looking for differences along isentropic trajectories covering different regions of the QCD phase diagram, passing far or close to the CEP or even intersecting the first-order critical line.}

\authorrunning{M. Motta et al.}
\titlerunning{Isentropic evolution of the matter in heavy-ion collisions}

\maketitle

\section{Introduction}
The goal of relativistic heavy-ion collisions is the exploration of the QCD phase diagram, looking for signatures of the transition from a system of colour-neutral hadrons to a deconfined plasma of quarks and gluons (QGP). Actually, in current ongoing experiments at RHIC and at the LHC, the transition occurs in the opposite direction: at very high energy, in the collisions of two nuclei, the amount of stopped baryonic matter is negligible, quarks and gluons are ``newly'' produced particles arising from the strong colour fields in the overlapping region. Quarks and gluons form a thermalized plasma undergoing an almost adiabatic expansion during which the latter cools down until reaching a temperature at which colour-singlet hadrons become again the active degrees of freedom.

First-principle lattice-QCD simulations show that, at vanishing baryon density (i.e. at baryo-chemical potential $\mu_{\rm B}\!=\!0$), the transition connecting the partonic and hadronic phases is actually a smooth crossover~\cite{Aoki:2006we}. This is the regime of relevance for the nuclear collisions at the LHC ($\sqrt{s_{\rm NN}}\!=\!2.76$ and 5.02 TeV) and at the highest center-of-mass energy at RHIC ($\sqrt{s_{\rm NN}}\!=\!200$ GeV) and this corresponds also to the regime at which the QCD transition occurred during the thermal history of the universe, around 1 $\mu$s after the Big Bang, when the temperature reached a value around 150-160 MeV~\cite{Borsanyi:2010bp}.

Unfortunately, due to the sign problem which prevents a Monte-Carlo sampling of the gauge-field configurations, lattice-QCD simulations cannot provide definite answers on the QCD thermodynamics and phase structure at finite baryon density, except for sufficiently small values of $\mu_{B}/T$ where, for instance, one can perform a Taylor expansion around $\mu_B\!=\!0$~\cite{Borsanyi:2012cr,Bazavov:2017dus}. Various effective models were then employed (see e.g. ~\cite{Klevansky:1992qe,Barducci:1993bh,Stephanov:1996ki}), suggesting that at large $\mu_{B}$ the transition should become of first order. Before turning into a crossover, such a first-order line in the $(\mu_{B},T)$-plane should end with a critical endpoint (CEP) in which the transition is of second order, characterized by an infinite correlation length. 

The study of the QCD phase-diagram in the region of non-vanishing baryon density (for a recent review see for instance~\cite{Ratti:2019tvj}) presents then several challenges. First of all one should provide a non-ambiguous, quantitative theoretical answer to the question concerning the existence and location of a CEP; so far lattice-QCD simulations tend to disfavour its presence for $\mu_B/T\lsim 2$~\cite{Bazavov:2017dus}. Secondly, one should suggest specific signatures of the QCD critical point to look for in the experimental data~\cite{Stephanov:2008qz,Stephanov:1999zu}. Finally, one should perform experiments able to explore the QCD matter in such a regime of high baryon-density. This can be achieved through the collisions of heavy nuclei at lower center-of-mass energies, characterized by a larger stopping of the incoming nucleons. For this purpose, experiments were performed in the past at AGS~\cite{Akiba:1996xf} and SPS~\cite{Alt:2007aa}, the Beam-Energy-Scan (BES) is currently ongoing at RHIC~\cite{Aggarwal:2010cw,Adamczyk:2017iwn}, a rich physics program is scheduled for the present and the near future at SPS~\cite{Melkumov:2019nja} and new infrastructures like NICA~\cite{Blaschke2016_1} and FAIR~\cite{Blaschke2016_2} are under construction. 
One of the difficulties in identifying unambiguous experimental signatures of the presence of a CEP and/or of the occurrence of a first-order transition in the hot QCD matter produced in relativistic heavy-ion collisions is that -- at variance with condensed-matter experiments -- one does not have the possibility to explore the phase diagram in a systematic way, tuning appropriate control parameters and performing measurements during the whole evolution of the system (except for the case of rare penetrating probes like photons and dileptons). One can only change the center-of-mass energy of the collisions and the nuclear species and possibly select events of different centrality; furthermore, one can only measure hadrons after their chemical and kinetic decoupling from the expanding fireball.

In heavy-ion collisions one produces a system which -- neglecting dissipative effects due to viscosity, heat conduction and charge diffusion -- undergoes an approximate isentropic expansion moved by pressure gradients along trajectories of constant entropy per baryon ($s/n_B\!=\!$ const), the higher the center-of-mass energy of the collision the higher the $s/n_B$ ratio. Hence, in performing theoretical calculations of experimental relevance, the quantities of physical interest must be evaluated along the above trajectories. This is what we plan to do in our paper, focusing in particular on two kind of thermodynamic quantities: the speed of sound and the generalized quark-number susceptibilities. Both of them have a deep experimental relevance. The squared speed of sound $c_s^2$ governs the response of the system to the initial energy-density gradients, leading to the collective acceleration of the fireball. If, during its evolution, the system undergoes a first-order transition during the mixed-phase the value of the speed of sound should be very low and this should affect the transverse-momentum distributions of the hadrons produced in the collision. The generalized susceptibilities, on the other hand, are related to the fluctuations of conserved charges (baryon number, electric charge and strangeness) which can be experimentally accessed through event-by-event measurements of the unbalance of protons and anti-protons~\cite{Adamczyk:2013dal}, opposite-charge particles~\cite{Adamczyk:2014fia} and, more recently, $K^+$ and $K^-$ mesons~\cite{Adamczyk:2017wsl}. They are expected to display huge oscillations and rapid changes of sign in the vicinity of a CEP, and one hopes that this will leave observables effects in the event-by-event measurements of identified particles produced at the chemical freeze-out, if the latter occurs sufficiently close to the CEP.

In our paper we evaluate the above quantities starting from two effective Lagrangians which display the same chiral-symmetry breaking/restoration pattern of QCD and include a simple modelling of quark confinement: the PNJL (Polyakov-Nambu-Jona-Lasinio) model~\cite{Ratti:2005jh} and the PQM (Polyakov-Quark-Meson) model~\cite{Mocsy:2003qw}.
We consider isentropic trajectories of relevance for the BES undergoing at RHIC and for future experiments at SPS, NICA and FAIR. The corresponding values of $s/n_B$ are estimated starting from the data provided by the STAR collaboration in~\cite{Adamczyk:2017iwn,Abelev:2008ab}. We study the values taken by the speed of sound and by the generalized susceptibilities during the isentropic evolution of the system until the estimated freeze-out point, looking for differences among cases in which the QCD transition occurs via a smooth crossover, a first-order transition or crossing a possible CEP.

Our paper is organized as follows. In Sec.~\ref{sec:models} we present the effective models employed for our calculations. In Sec.~\ref{sec:results} we display our results for the phase diagram of the two models and for the evolution of the speed of sound and of the generalized quark susceptibilities along very different isentropic trajectories. Finally, in Sec.~\ref{sec:conclusions} we discuss the obtained results and their possible experimental relevance, providing also some perspective for future developments.

%%%%%%%%%%%%%%%%%%%%%%%%%%%%%%%%%%%%%%%%%%%%%%%%%%%%%%%%%%%%
\section{The PNJL and PQM models}\label{sec:models}
In this section we present the two effective Lagrangians used in our study: the PNJL and the PQM model. The two models share some similarities, in particular in the effective implementation of quark confinement, but they differ in the way in which the interaction among quarks is described: via a Fermi-like four-fermion vertex in the PNJL model, via the exchange of scalar/pseudoscalar mesons representing bosonized quark fields in the PQM model. Both models have been already described at length in the literature and for more details we refer the interested reader to the original publications on the subject~\cite{Ratti:2005jh,Mocsy:2003qw}. 

\subsection{The Polyakov field and the effective implementation of confinement}
In QCD -- for infinite quark mass -- the order parameter of the confinement/deconfinement phase transition is the Polyakov loop, expressed in terms of the Euclidean temporal component of the gauge field $A_4^E(\tau)\equiv -iA_0(t=-i\tau)$ as follows
\begin{equation}
    \Phi\!\slr{A_\mu}\equiv\frac{1}{N_c}{\mrm Tr}L_{\vec x}=\frac{1}{N_c}\mrm{Tr}\bigg\{\mathcal{P}\mrm{exp}\Big[\,i\!\int_0^\beta\! d\tau A_4^E(\tau,\vec x)\Big]\bigg\}.
    \label{eq:PF}
\end{equation}
Its expectation value is related via the equation
\beq
\langle\Phi\rangle=e^{-\beta\Delta F_Q}
\eeq
to the change in free-energy occurring adding an isolated heavy quark into  the system: it vanishes in the confined phase, where adding an isolated colour charge requires an infinite amount of energy; it is non-zero in the deconfined phase, where this suppression of coloured states is absent.
From the mathematical point of view $\langle\Phi\rangle\ne 0$ leads to the spontaneous breaking of the global $Z_{N_c}$ symmetry, related to the transformations
\beq
L_{\vec x}\to z^k L_{\vec x}\quad{\rm with}\quad z^k=e^{i2\pi k/N_c}
\eeq
which multiplies all the Polyakov lines $L_{\vec x}$ by the same phase factor $z^k$. In the pure $SU(N_c)$ gauge theory this is an exact symmetry of the Yang Mills action. Light dynamical quarks introduce an explicit breaking of the $Z_{N_c}$ symmetry ($N_c=3$ in QCD), nevertheless the Polyakov loop, in light of its physical meaning, remains a useful quantity to identify the different phases predicted by the theory.

Both in the PNJL and in the PQM model the Polyakov loop $\Phi$ (and $\Phib$ for the antiquark sector) plays the role of a constant background field whose value is determined looking for the stationary points of the thermodynamic potential and whose role is to suppress the contribution of free quarks to the thermodynamic quantities. In order to ensure the occurrence of a phase transition, an effective temperature-dependent potential for the $\Phi$ and $\Phib$ fields is introduced into the Lagrangian, with the constraint of respecting the $Z_3$ symmetry of the theory. In this paper we choose it of the following polynomial form:
\begin{equation}
%    \begin{split}
        \mathcal{U}_{\rm Pol}\left(\Phi,\Phib;T\right)=T^4\bigg\{-\frac{b_2(T)}{2}\bar\Phi\Phi-\frac{b_3}{6}(\Phi^3+\bar\Phi^3)+\frac{b_4}{4}(\bar\Phi\Phi)^2\bigg\}.\label{eq:polynomial}
%    \end{split}
\end{equation}
The function $b_2(T)$ is parametrized as follow
\begin{equation}
    b_2(T)=a_0+a_1\Big(\frac{T_0}{T}\Big)+a_2\Big(\frac{T_0}{T}\Big)^2+a_3\Big(\frac{T_0}{T}\Big)^3
\end{equation}
and the coefficients of the potential are fitted in order to reproduce the lattice data for pure Yang-Mills Theory ~\cite{Ratti:2005jh}: their values are given Table \ref{tab:ULog}. Concerning $T_0$, in the case of 2+1 light dynamical quarks addressed in our paper, we set it to 182 MeV. 
\begin{table}[h]
    \centering
    $
    \begin{array}{cccccc}
         \toprule
         a_0 &  a_1 & a_2 & a_3 & b_3 & b_4 \\
         \midrule
         6.75\quad &   -1.95\quad &  2.625\quad &   -7.44\quad &  0.75\quad & 7.5\\
         \bottomrule
    \end{array}
    $
    \caption{Parameters of the Polyakov-loop potential}    
         \label{tab:ULog}    
\end{table}

\subsection{The PNJL and PQM Lagrangians}\label{subsec:PNJL}
The PNJL and PQM Lagrangians can be written as ${\cal L}={\cal L}_0+\Delta{\cal L}$, where ${\cal L}_0$ is the piece common to both models (including the quark kinetic term, the coupling with the background Polyakov field and the Polyakov potential), while $\Delta{\cal L}$ describes the quark interaction, either through a contact term (in the PNJL model) or mediated by light scalar/pseudoscalar mesons (PQM model). We start introducing the common piece of the Lagrangian, which reads
\begin{equation}
	\mc{L}_0 \equiv\bar \psi\left[i\gamma_{\mu}\!\left(D^{\mu}\!-i\delta^{\mu 0}\widehat{\mu}_f\right)\right]\psi-\,\mathcal{U}_{\rm Pol}\left(\Phi,\Phib;T\right).
\label{eq:LagrangianPNJLPQM1}
\end{equation}
%\end{multline}
In the Dirac term the covariant derivative is defined as $D_\mu\equiv\partial_\mu-iA_\mu$, where one sets $A_\mu\equiv\delta_{\mu 0}A_0$. The strong
coupling constant $g_s$ is absorbed in the definition of $A_\mu(x)\equiv g_s \mc{A}_\mu^a(x)\frac{\lambda^a}{2}$, where $\mc{A}_\mu^a(x)$ is the $SU_c(3)$ gauge field and $\lambda^a,\ a=1,..,8$ are the Gell-Mann matrices of the $SU_c(3)$ group. 
The symbol $\widehat{\mu}_f$ indicates the chemical-potential matrix associated to the various quark flavours,   $\widehat{\mu}_f\equiv\mrm{diag}(\mu_u,\mu_d,\mu_s)$, which are conserved by strong interactions. Notice that the structure of the Lagrangian entails that the background gauge field enters into the quark propagator as a shift of the chemical potential, $\mu\longrightarrow\mu+A_0\equiv\mu+iA_4^E$.
Finally, the term $\mathcal{U}_{\rm Pol}$ is the Polyakov-loop potential given in Eq.~(\ref{eq:polynomial}).

It is useful to remind the relation between the quark chemical-potentials and the ones associated to the conserved QCD charges at the hadronic level, which are the ones of actual experimental relevance:
\begin{equation}
    \begin{cases}
    \mu_u=\frac{1}{3}\mu_B+\frac{2}{3}\mu_Q\\
    \mu_d=\frac{1}{3}\mu_B-\frac{1}{3}\mu_Q\\
    \mu_s=\frac{1}{3}\mu_B-\frac{1}{3}\mu_Q-\mu_S\;.
    \end{cases}
\end{equation}
Here $\mu_B, \mu_Q, \mu_S$ are the baryon, electric-charge and strangeness chemical potentials.

We now discuss the interaction terms of the Lagrangian, in which the two models differ.

\begin{table}[h!]
    \centering
    $
    \begin{array}{ccccc}
         \toprule
         G &  K & \Lambda\text{[MeV]} & m_{\ell0}\text{[MeV]} & m_{s0}\text{[MeV]}\\
         \midrule
         3.67/\Lambda^2\quad &   -12.36/\Lambda^5 \quad &  602.3\quad & 5.5\quad & 140.7\quad \\
         \bottomrule
    \end{array}
    $
    \caption{Parameters of the NJL model}    
         \label{tab:NJLParameters}    
\end{table}

We start from the PNJL model, for which we have:
\begin{equation}
\begin{split}
\Delta\mc{L}_\mrm{PNJL}=&-\bar\psi\widehat m_0\psi+ \frac{1}{2}G\sum_{a=0}^8[(\bar{\psi}\lambda^a \psi)^2+(\bar{\psi}i\gamma^5\lambda^a\psi)^2]+\\
&+K\{\text{det}[\bar{\psi}(1+\gamma^5)\psi]+\text{det}[\bar{\psi}(1-\gamma^5)\psi]\}.
\end{split}
\label{eq:LagrangianPNJL}
\end{equation}
In the above $\widehat{m}_0$ is the diagonal mass matrix for the quarks, $\widehat{m}_0\equiv\mrm{diag}(m_{0u},m_{0d},m_{0s})$,  $\lambda^a\;(a\!=\!1,2,...,8)$ are the $3\times 3$ Gell-Mann matrices of the $SU_f(3)$ group and  $\lambda^0=\sqrt{2/3}\mathbb{1}$.
$G$ and $K$ are the coupling constant for the 4-fermion vertex and 6-fermion vertex, with dimensions $[E]^{-2}$ and $[E]^{-5}$, respectively.
The six-fermion term is know as 't-Hooft determinant and is introduced to explicitly break the axial $U_A(1)$ symmetry, which is not a symmetry of QCD at the quantum level, as confirmed by the high mass of the pseudoscalar $\eta'$ meson.
Due to the above interaction terms the PNJL model in 4 space-time dimensions is not renormalizable and it has to be considered just a low-energy effective theory. To regularize the divergent integrals one has to introduce a momentum cutoff $\Lambda$, representing a further free parameter of the model together with the couplings $G$ and $K$ and the quark masses: they are all summarized in Table \ref{tab:NJLParameters}. We work in the isospin-symmetric case, with $m_{0u}=m_{0d}\equiv m_{0\ell}$, setting also $\mu_u=\mu_d\equiv \mu_\ell$.

\begin{table*}[!htb]
	\begin{center}	
		\begin{tabular}{lccccccc}
			\toprule
                Constant & $f_\pi$ & $m_\pi$ & $m_\mrm{K}$ & $m_\eta^2 + m_{\eta'}^2$ & $m_\sigma$ & $m_\ell$ & $m_\mrm{s}$\\
			\midrule
				Value [MeV] & 92.4 & 135 & 497.7 & $514.8^2 + 957.8^2$ & 728.9 & 367.7 & 549.5\\
			\bottomrule
		\end{tabular}
	\end{center} 
\caption{Values of the decay constants of pseudoscalar mesons, of the meson masses and of the constituent mass of light (up and down) and strange quarks in the vacuum to which the parameters of the PQM model are adjusted to. These values are the same as in the PNJL model calculation.}	
	\label{tab:chiral_pot_constants_PNJL}
\end{table*}

Considering now the PQM model, the terms of the Lagrangian describing the scalar and pseudoscalar mesons and their interaction with the quark field read
\begin{multline}
	\Delta\mc{L}_\mrm{PQM}= - g \, \bar{\psi} \slr{ \sum_{a=0}^8 \lambda_a \blr{\sigma_a + i\, \gamma_5 \pi_a} } \psi \,+\\
	+ k \blr{\det\phi + \det\phi^\dagger} + \Tr \left[ H \blr{\phi + \phi^\dagger} \right] -\\
	- m^2\, \Tr \blr{\phi^\dagger\phi} - \lambda_1 \left[ \Tr\blr{\phi^\dagger\phi} \right]^2 - \lambda_2\, \Tr \blr{\phi^\dagger\phi}^2 + \\
	+ \Tr\blr{\pd_\mu \phi^\dagger \,\pd^\mu\phi} \;,
	\label{eq:lagrQM}
\end{multline}
%e%e%e%
where $\phi$ combines the scalar and pseudoscalar meson fields
%e%e%e%
\be
	\phi = \sum_{a=0}^8 T_a\, \phi_a = \sum_{a=0}^8 \frac{\lambda_a}{2} \blr{\sigma_a + i\, \pi_a}\;,
	\label{eq:mesonMatrix}
\ee
%e%e%e%
$g$ denotes the Yukawa coupling between quarks and mesons,
$k$ is the coefficient of the $U_A(1)$ symmetry breaking term and  $H=T_a\,h_a$. The coefficients $k$, $h_a$, $m^2$, $\lambda_1$ and $\lambda_2$
are the parameters of the mesonic Mexican-hat potential  \cite{Levy:1967,Lenaghan:2000ey,Schaefer:2009ui}
which describes the spontaneous breaking of chiral symmetry, giving the meson fields their expectation values $\sbar_a$. The explicit breaking of chiral symmetry in the PQM model arises from the tilt of the mesonic potential provided by the term $H(\phi+\phi^\dagger)$, playing the role of the $-\bar\psi\widehat m_0\psi$ term in the PNJL Lagrangian. The quantities used to fix the parameters of the PQM model are summarized in Table~\ref{tab:chiral_pot_constants_PNJL}.

\subsection{The mean-field approximation and the thermodynamic potential }
In this section we display the equations allowing one to study the thermodynamics of the PNJL and PQM models in the mean field approximation, at the basis of the numerical results presented  in Sec.~\ref{sec:results}.
Within this setup a system of strongly-coupled quarks is described as a collection of non-interacting quasi-particles, endowed with effective masses obtained through the minimization of the resulting thermodynamic potential.

In the two models the effective quark mass has a different origin. In the PQM case it arises from the Yukawa coupling with the meson fields, which develop a non-vanishing expectation value, leading to $m_\ell= g \sbar_\ell $ and $m_\s = \sqrt{2} g \sbar_\s$ for the light and strange quarks, respectively~\cite{Schaefer:2008hk}.
In the PNJL case it arises from the quark self-interaction and is obtained linearizing the Lagrangian given in Eq.~(\ref{eq:LagrangianPNJL}) with respect to the fluctuations of the composite operator $\bar\psi_f\psi_f=\langle\bar\psi_f\psi_f\rangle+\delta_f\equiv\varphi_f+\delta_f$. One obtains:
\begin{multline}
\Delta\mc{L}_\mrm{PNJL}^{\rm MF}=-\bar\psi\widehat m_0\psi+\sum_{i\ne j,k}\bar\psi_i\left[4G\varphi_i+2K\varphi_j\varphi_k\right]\psi_i\\
-2G\sum_i\varphi_i^2-4K\varphi_u\varphi_d\varphi_s
\end{multline}
and hence the effective quark masses (we assume exact isospin symmetry, setting $m_u=m_d\equiv m_\ell$) turn out to be expressed in terms of the chiral condensates $\varphi_f\equiv\langle\bar\psi_f\psi_f\rangle$ of the various quark flavours 
\begin{equation}
\begin{cases}
    m_\ell=m_{0\ell}-4G\varphi_\ell-2K\varphi_\ell\varphi_s\\
     m_s=m_{0s}-4G\varphi_s-2K\varphi^2_\ell.
     \end{cases}
    \label{eq:massgap}
\end{equation}
Getting the quark masses requires the self-consistent solution of the above system of coupled gap equations, where the mass of quarks of flavour $i$ depends on all chiral condensates
\begin{equation}
\begin{split}
\varphi_i=-2N_c\int^\Lambda\frac{d^3p}{(2\pi)^3}&\frac{m_i}{E_p^i}[1-f_q(\Phi,\bar\Phi,E_p^i;T,\mu_i)-\\&-f_{\bar q}(\Phi,\bar\Phi,E_p^i;T,\mu_i)].\label{eq:gapeq1}
\end{split}
\end{equation}
Here $E_p^i\!=\!\sqrt{p^2+m_i^2}$ is the energy of the dressed quark of flavour $i$, and the modified Fermi distributions are given by
  \begin{multline}
      f_q(\Phi,\bar\Phi,E_p^i;T,\mu_{i})\equiv\\\frac{\Phi\e^{-\beta(E_p^i-\mu_i)}+2\bar\Phi \e^{-\beta(E_p^i-\mu_i)}+\e^{-3\beta(E_p^i-\mu_i)}}{1+3\Phi\e^{-\beta(E_p^i-\mu_i)}+3\bar\Phi\e^{-2\beta(E_p^i-\mu_i)}+e^{-3\beta(E_p^i-\mu_i)}}
  \end{multline}
  and
\begin{equation}
    f_{\bar q}(\Phi,\bar\Phi,E_p^i;T,\mu_{i})=f_q(\bar\Phi,\Phi,E_p^i;T,-\mu_{i}).
\end{equation}

Also the thermodynamic potential can be conveniently written as $\Omega\equiv\Omega_0+\Delta\Omega$, where $\Omega_0$ is the part common to both models, while $\Delta\Omega$ includes the model-dependent terms. At the mean field level one has
\begin{multline}
\Omega_0^{\rm MF}/V=\mathcal{U}_{\rm Pol}(\Phi,\Phib;T)-2\Nc\sum_{f}\int\limits^\Lambda\!\frac{\mrm{d^3}p}{(2\pi)^3}\,E_p^f\\
-2T\sum_f\!\int\limits^\infty\!\frac{d^3p}{(2\pi)^3}\slr{z_{q}\!\blr{\Phi,\Phib,E_p^f;T,\mu_f}+z_{\bar q}\!\blr{\Phi,\Phib,E_p^f;T,\mu_f}},\label{eq:OmegaMF_common}
\end{multline}
having defined
\beq
    z_q\equiv\mrm{ln}[1+3\Phi\e^{-\beta(E_p^f-\mu_f)}+3\Phib\e^{-2\beta(E_p^f-\mu_f)}+\e^{-3\beta(E_p^f-\mu_f)}]\label{eq:effective-q}
\eeq
and
\beq
    z_{\bar q}\equiv\mrm{ln}[1+3\Phib\e^{-\beta(E_p^f+\mu_f)}+3\Phi\e^{-2\beta(E_p^f+\mu_f)}+\e^{-3\beta(E_p^f+\mu_f)}].\label{eq:effective-qbar}
\eeq
Concerning the model-dependent contribution to the thermodynamic potential, in the PNJL case one has:
\beq
\Delta\Omega_{\rm PNJL}^{\rm MF}/V=2G(2\varphi_\ell^2+\varphi_s^2)+4K\varphi_\ell^2\varphi_s.
\eeq
For the PQM model the contribution from the expectation value of the meson fields reads:
\begin{multline}
	\Delta\Omega_{\rm PQM}^{\rm MF}/V = - \frac{k}{2\sqrt{2}} \sbar_\ell^2 \sbar_\s 
	 - h_\ell\sbar_\ell - h_\mrm{s}\sbar_\s + \\
	 + \frac{m^2}{2} \blr{\sbar_\ell^2+\sbar_\s^2} + \\
	 + \frac{ \lone +  {\ltwo}/{2} }{4} \, \sbar_\ell^4 + \frac{ \lone +  \ltwo }{4} \,  \sbar_\s^4 + \frac{\lone}{2} \, \sbar_\ell^2 \sbar_\s^2.
	\label{eq:LSMPot}
\end{multline}
At the mean field level the thermodynamic potential is a function of the effective quark masses and of the expectation value of the Polyakov fields,
\beq
\Omega^{\rm MF}=\Omega^{\rm MF}(m_\ell,m_s,\Phi,\Phib),\nonumber
\eeq
which have to be self-consistently determined requiring $\Omega^{\rm MF}$ to be stationary under variation of the above quantities, i.e.
\beq
\frac{\partial \Omega^{\rm MF}}{\partial m_\ell}=
\frac{\partial \Omega^{\rm MF}}{\partial m_s}=\frac{\partial \Omega^{\rm MF}}{\partial\Phi}=\frac{\partial \Omega^{\rm MF}}{\partial\Phib}=0.
\eeq
The first two conditions lead to the mass-gap equations (\ref{eq:gapeq1}), which in the PNJL model were independently obtained expressing the chiral condensates $\varphi_i$ in terms of the Hartree quark propagators. Equivalently, the dependence on the effective quark masses $m_\ell$ and $m_s$ can be traded for the one on the chiral condensates $\varphi_\ell$ and $\varphi_s$ or on the expectation values of the meson fields $\sigma_\ell$ and $\sigma_s$.
Notice that in the present MF approximation, replacing the meson fields by their expectation values, the associated kinetic contribution in the PQM Lagrangian in Eq.~(\ref{eq:lagrQM}) vanishes.

A final comment on the UV behaviour of the two models is in order. The second term in the RHS of Eq.~(\ref{eq:OmegaMF_common}) contains the sum of the (negative) zero-point energies of the various fermionic modes. It is divergent and is regularized by the UV cutoff $\Lambda$. In the PNJL case the results depend explicitly on this cutoff, which -- as already mentioned -- has to be viewed as a parameter of the model, fixed by matching some zero-temperature/density observables. On the other hand, the PQM model is renormalizable and in principle one could cancel the dependence on $\Lambda$ of the vacuum fluctuations with the dependence on the cutoff of the parameters of the meson potential in Eq.~(\ref{eq:LSMPot}), so that the final physical results do not depend on the choice of the UV regulator.
Here however, in order to perform a consistent comparison, we simply regularize the UV divegence through the same ultraviolet cutoff $\Lambda$ employed in the PNJL model. 

%%%%%%%%%%%%%%%%%%%%%%%%%%%%%%%%%%%%%%%%%%%%%%%%%%%%%%%%%%%%%%%%%

\section{Results}\label{sec:results}
Our results focus on the phase-diagram of the PNJL and PQM models, on the nature of the chiral-deconfinement transition, on the location of the critical endpoint (CEP), on the speed of sound and on the generalized susceptibilities, the latter being of interest since associated to the thermal fluctuations of conserved charges. All our numerical calculations are performed along isentropic trajectories. The comparison of the angular distributions of soft identified hadrons with the results of hydrodynamic calculations suggests in fact a very low value of the viscosity to entropy-density ratio, compatible with the conjectured universal lower bound $\eta/s=1/(4\pi)$ predicted by the gauge-gravity duality. Dissipative effects are then small in the fireball produced in heavy-ion collisions, which evolves along trajectories of constant entropy per baryon: both the entropy and the baryon density get dilute due to the expansion of the system (which has a positive pressure with respect to the surrounding vacuum), but their ratio $s/n_B$ remains constant within each fluid-cell.

In order to provide results of phenomenological relevance we need to estimate both the entropy-per-baryon ratio $S/B$ and the initial entropy density of the system arising from the heavy-ion collisions at the various nucleon-nucleon center-of-mass energies explored in the BES at RHIC, from 7.7 GeV to 200 GeV. We start estimating the initial entropy density $s_0$, assuming that its value is proportional to the measured rapidity density of charged particles $dN^{\rm ch}/d\eta$. For Au-Au collisions at $\sqrt{s_{\rm NN}}=200$ GeV the value $s_0=84\,{\rm fm}^{-3}$, once inserted in the initial condition of hydrodynamic calculations, was shown to satisfactory reproduce soft-hadron distributions~\cite{Luzum:2008cw} and, later on, it was widely employed in the literature, e.g.~\cite{Alberico:2011zy,Beraudo:2014boa}. Taking $\tau_0=1$ fm/c as an estimate of the initial thermalization time and integrating over the transverse plane the profile provided by a Glauber calculation one gets $dS/dy\approx 4700$ for the entropy per unit rapidity in central Au-Au collisions; taking into account that, for a pion gas around the chemical freeze-out temperature, one has $S\gsim 4 N$, this compares well with the observed rapidity density of charged particles~\cite{Back:2003ff}.
The initial entropy density $s_0$ at lower center-of-mass energies is obtained rescaling the estimate at $\sqrt{s_{\rm NN}}=200$ GeV according to the lower values of $dN^{\rm ch}/dy$~\cite{Adamczyk:2017iwn,Abelev:2008ab}. Also the $S/B$ ratio at the various center-of-mass energies is estimated from the yields of identified hadrons -- $\pi^\pm$, $K^\pm$, $p/\overline p$ -- quoted in Refs.~\cite{Adamczyk:2017iwn,Abelev:2008ab}, still assuming that each particle carries about 4 unit of entropy. Our results for $s_0$ and $S/B$ are collected in Table~\ref{tab:ise}, where we also quote the values of the kinetic freeze-out temperatures obtained in~\cite{Adamczyk:2017iwn,Abelev:2008ab} through a blast-wave fit of the transverse-momentum distributions of identified hadrons.

\begin{table}[!t]
    \begin{center}
    \begin{tabular}{cccc}
         \toprule
            $\sqrt{s_{\rm NN}}$ [GeV]\qquad &  $s_0$ [fm$^{-3}$] \qquad & $S/B$ \qquad&$T^{\rm fo}_{\rm kin}$\\
         \midrule
         7.7 &   29.6  &  17.5&116 \\
         11.5 & 35.3 & 26.7&118\\
         19.6 & 43.0 & 45.8&113\\
         27.0 & 45.8 & 56.8&117\\
         39.0 & 47.6& 84.3&117\\
         62.4 & 60.2& 123.9&99\\
         130  & 70  & 277.6&98\\
         200 & 84 & 331.6&89\\
         \bottomrule
    \end{tabular}
    \caption{Estimate of the initial entropy density and of the entropy per baryon in Au-Au collisions at different center-of-mass energies. We also quote the kinetic freeze-out temperature obtained in Refs.~\cite{Adamczyk:2017iwn,Abelev:2008ab} through a blast-wave-fit.}\label{tab:ise}
\end{center}
\end{table}

\subsection{Phase diagram and isentropic trajectories}
\begin{figure*}[!ht]
	\centering
	\includegraphics[width=.49\textwidth]{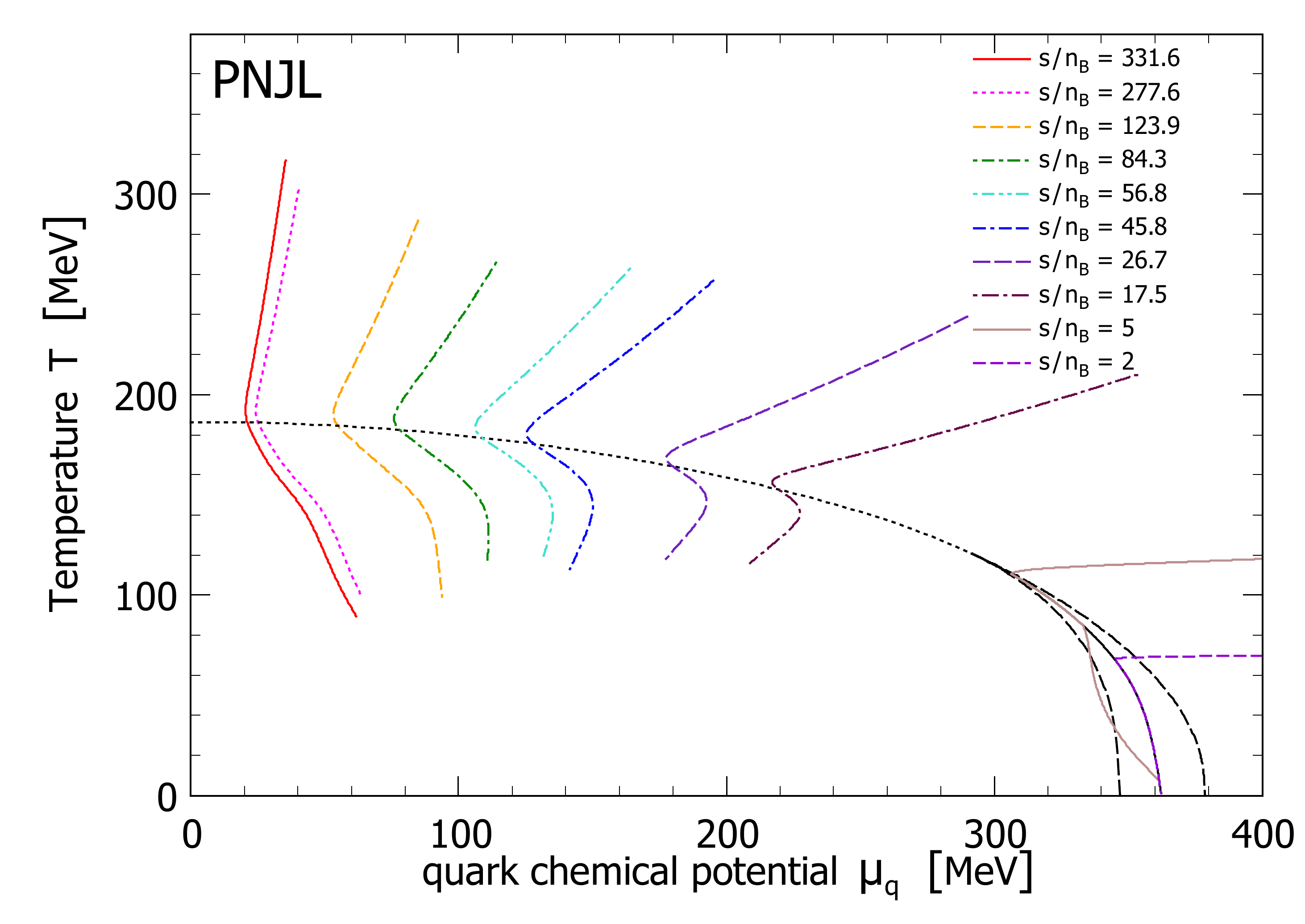}
		\hfill
	\includegraphics[width=.49\textwidth]{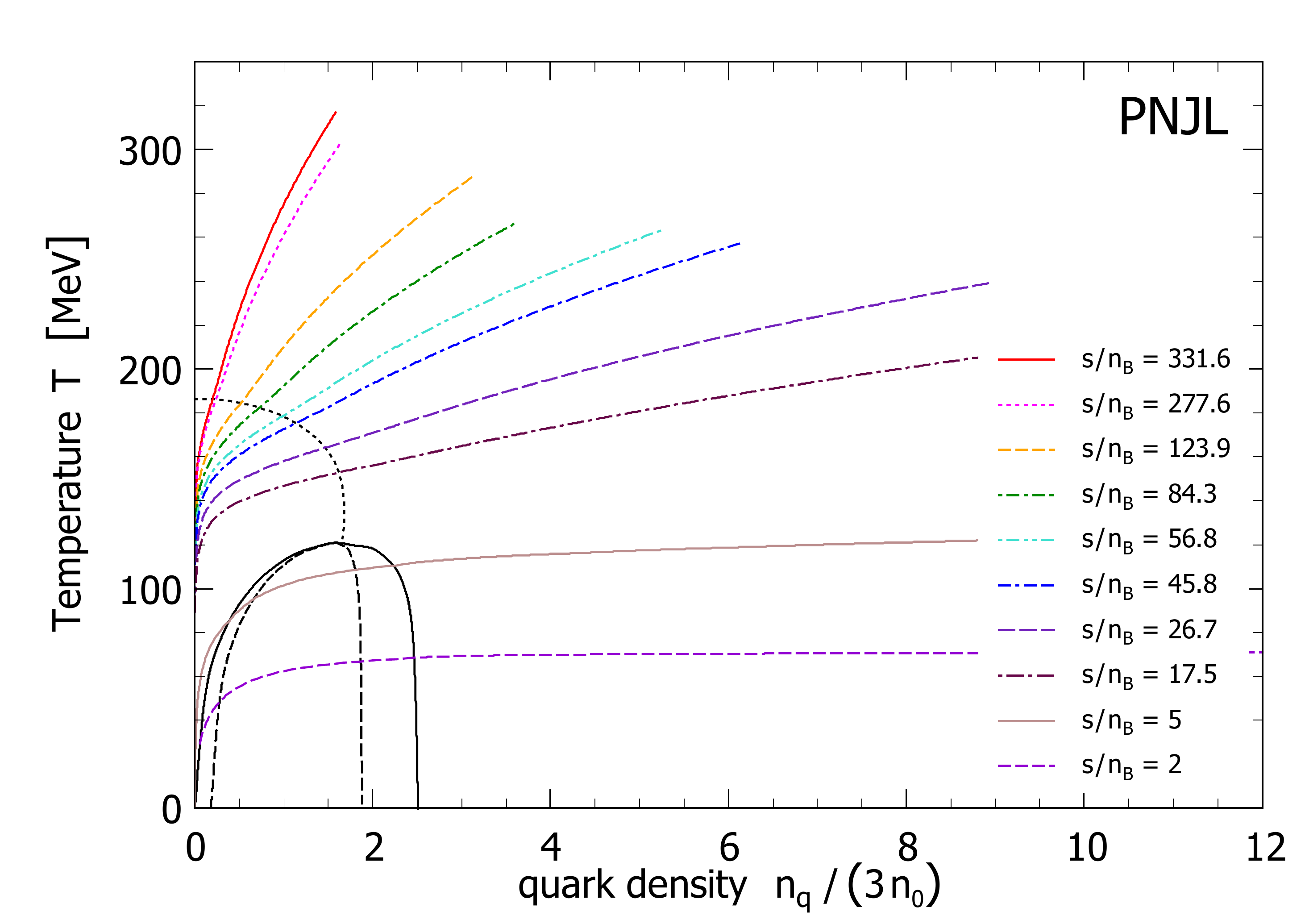}
	\caption{Phase diagram of the PNJL model expressed in terms of the chemical potential (left) and of the baryon density (right). The black, dotted lines represent the chiral crossover, identified through the inflection point of the effective mass of light quarks. The continuous, black curves represent the first-order transition lines. The black, dashed spinodal lines represent the boundaries of the regions where a homogeneous metastable phase is possible. We also show, in different colours, the isentropic trajectories referring to the $s/n_B$  values given in Table~\ref{tab:ise}. Each curve starts at a temperature $T_0$ corresponding to the values of $s_0$ given in Table~\ref{tab:ise} and is plotted down to the corresponding kinetic freeze-out temperature $T^{\rm fo}$. We also plot two isentropic trajectories so far not covered by the BES and entering into the first-order region.}
	\label{fig:PNJL_PD_PDn}
\end{figure*}
%m%m%m%
%f%f%f%
\begin{figure*}[!ht]
	\centering
	\includegraphics[width=.49\textwidth]{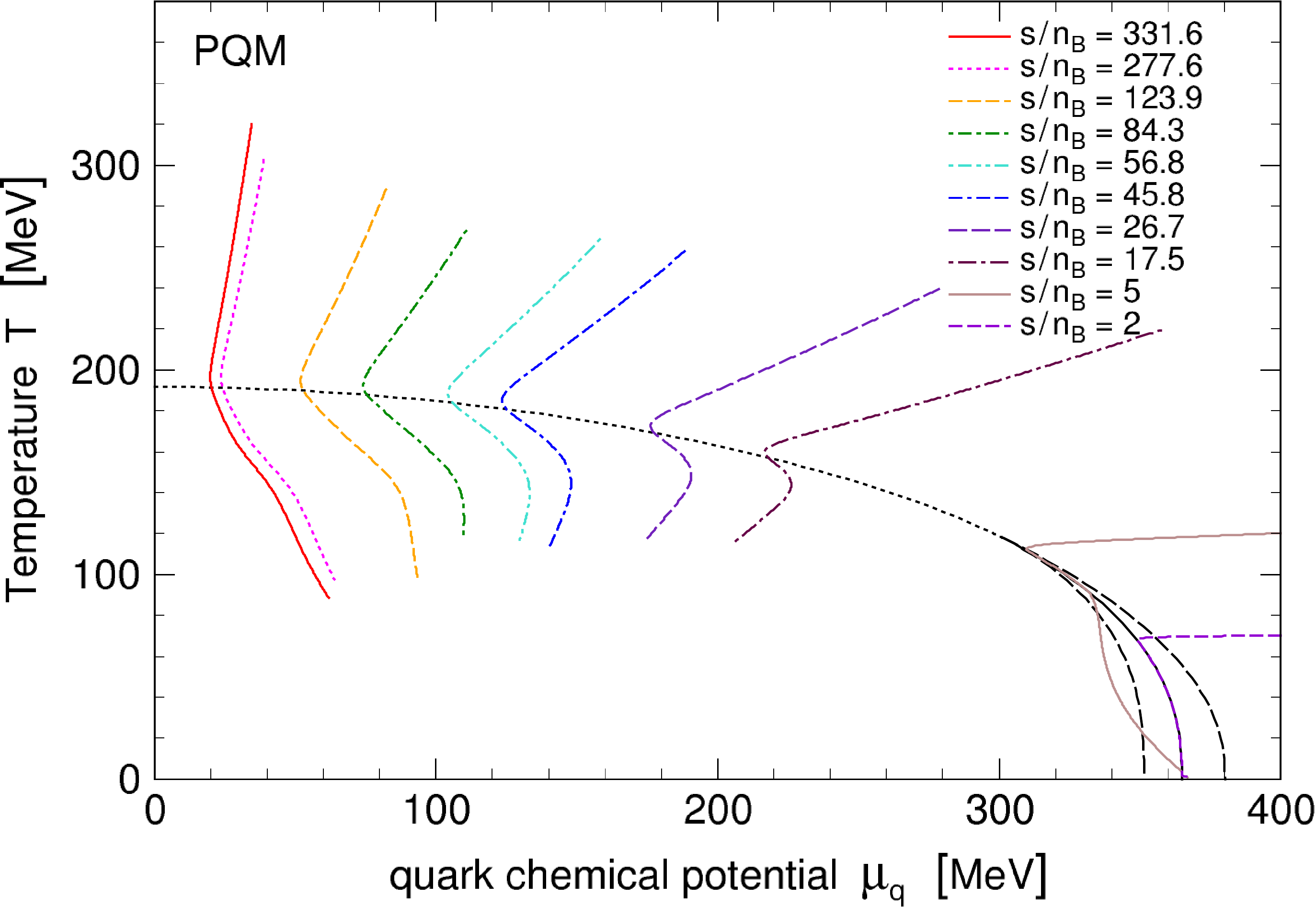}
	\hfill
	\includegraphics[width=.49\textwidth]{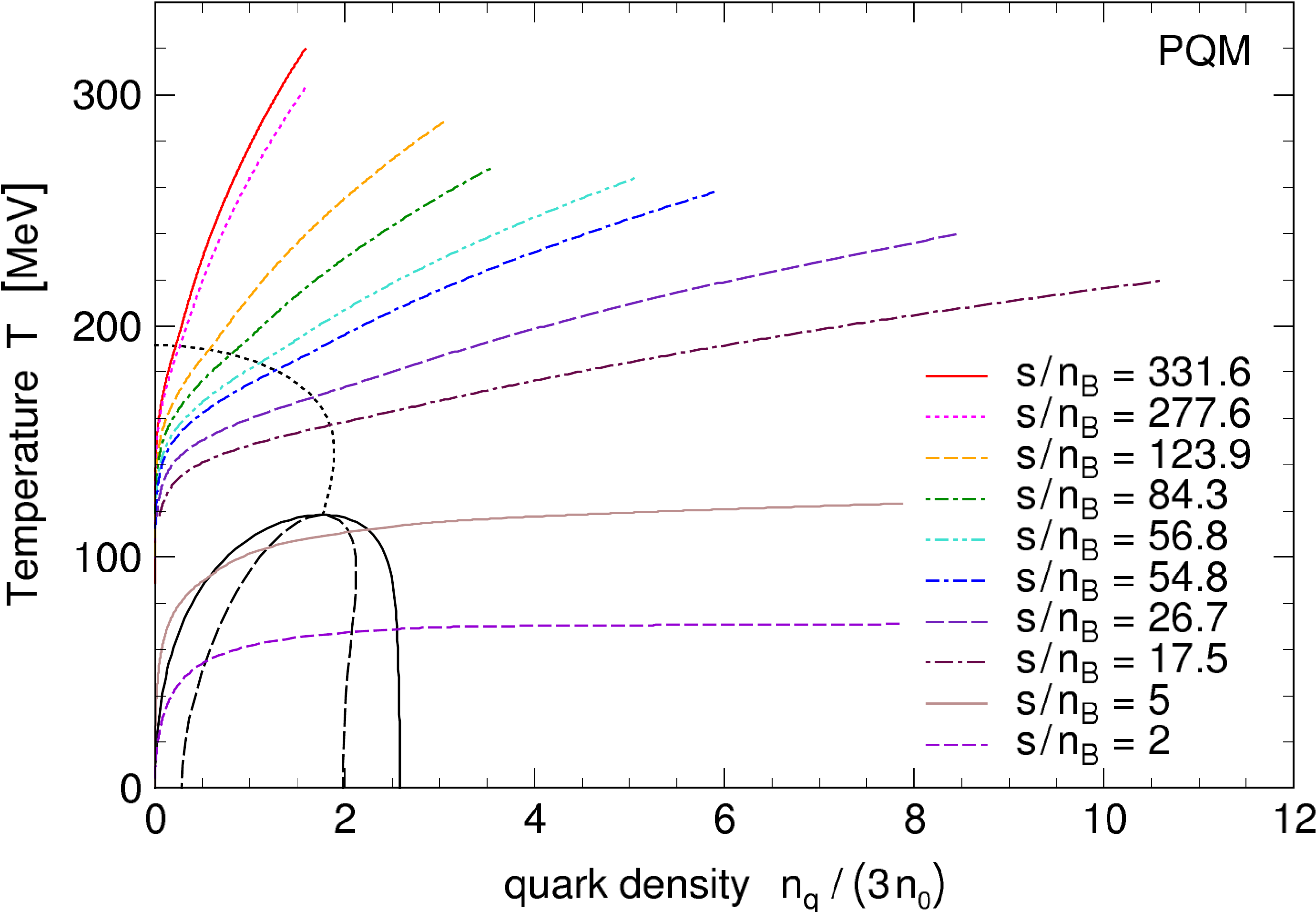}
	\caption[]{The same as in Fig.~\ref{fig:PNJL_PD_PDn}, but for the PQM model.}
	\label{fig:PQM_PD_PDn}
\end{figure*}
%f%f%f%
%f%f%f%
We start our analysis with the study of the phase diagram of the two models and of the evolution of the system along isentropic trajectories corresponding to values of $s/n_B$ of interest for the ongoing BES at RHIC. We display our results both in the $\mu_q\!-\!T$ and in the $n_q\!-\!T$ planes, the last case allowing one to get a deeper physical insight when the system, during its evolution, meets a first-order phase transition. Our findings are shown in Figs.~\ref{fig:PNJL_PD_PDn} and~\ref{fig:PQM_PD_PDn}. There is a wide region in the phase diagram in which the transition from the chirally-symmetric, deconfined phase to the chirally-broken, confined one is actually a crossover. In this case there is no well-defined location of the transition associated to an unambiguous order parameter, but there are several quantities displaying peaks in a quite narrow range of temperatures and chemical potentials. We decide to identify the crossover between the two phases -- shown as a black, dotted line in Figs.~(\ref{fig:PNJL_PD_PDn}) and~(\ref{fig:PQM_PD_PDn}) -- through the inflection point of the effective mass of the light quarks. We also display the isentropic trajectories followed by the system corresponding to the $s/n_B$ values quoted in Table~\ref{tab:ise}; for each case, they are plotted starting from $s_0$ down to the estimated kinetic freeze-out temperature $T_{\rm kin}^{\rm fo}$. In both the PNJL and the PQM models, for the currently accessible values of $s/n_B$, the transition of the system between the two phases occurs in the crossover region. 

%f%f%f%
\begin{figure*}[!ht]
	\centering
	\includegraphics[width=.49\textwidth]{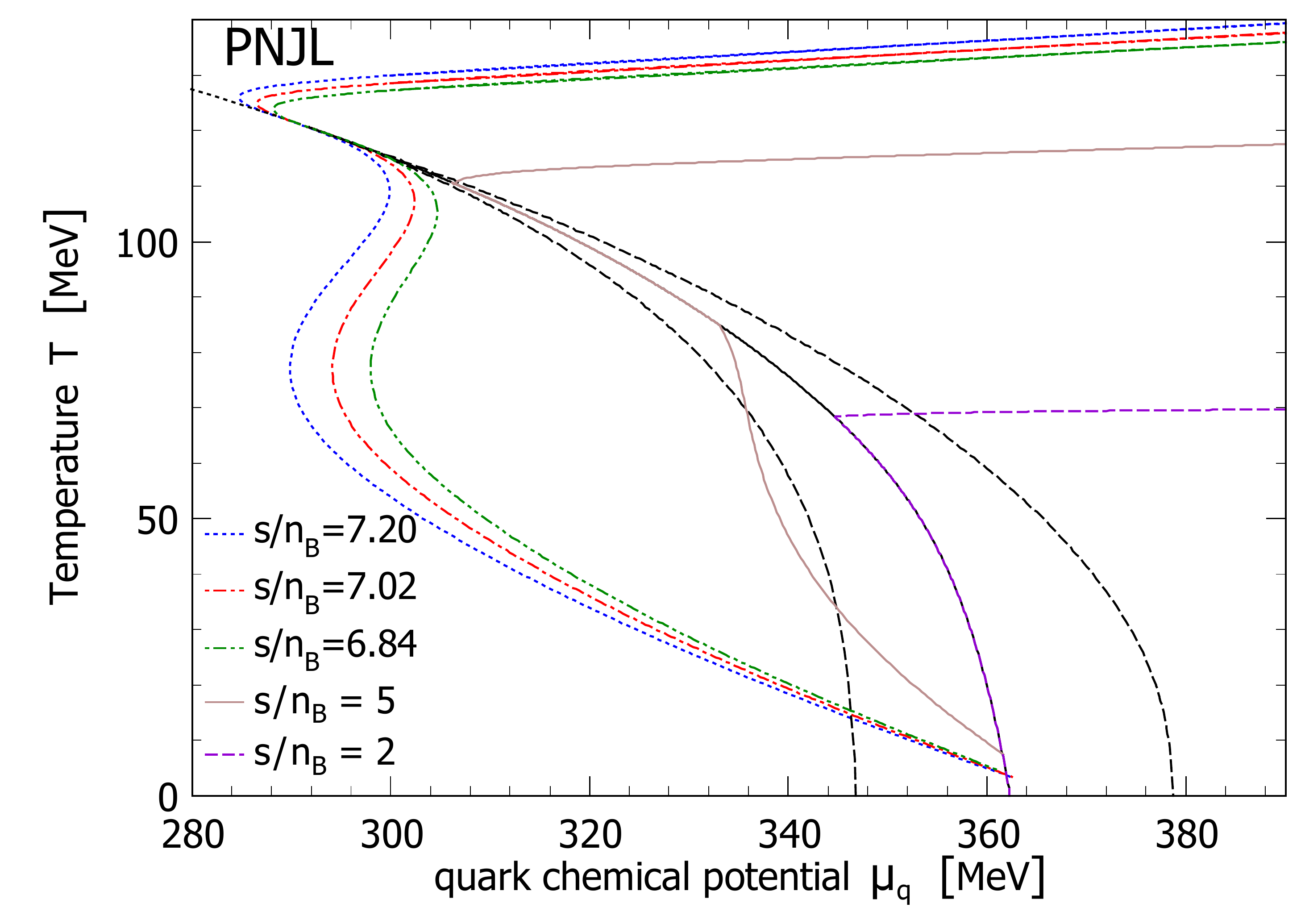}	
	\hfill
	\includegraphics[width=.49\textwidth]{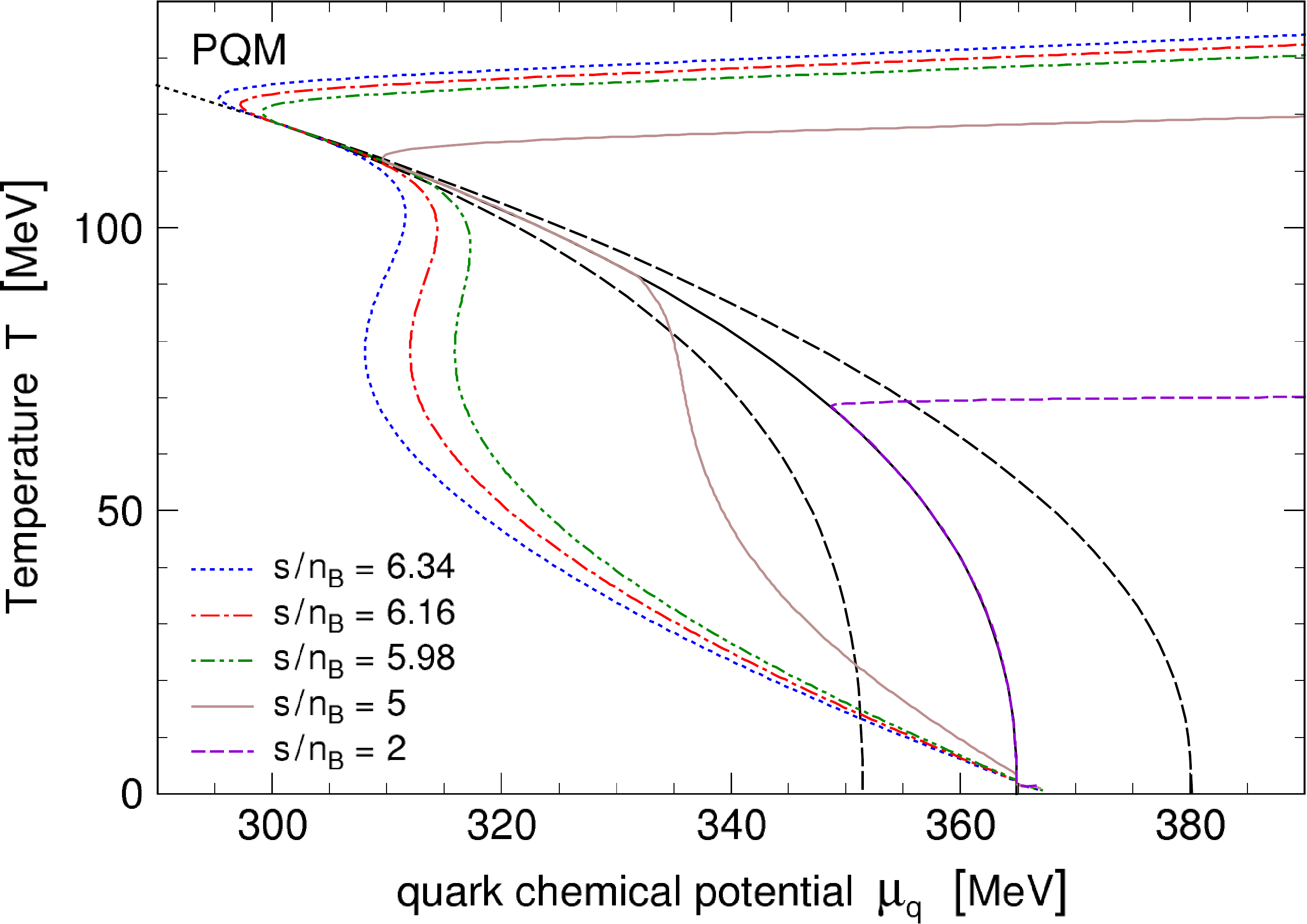}
	\caption{Zoom of the first-order region in the phase-diagram of the PNJL (left panel) and PQM (right panel) model. The black, dashed lines are the boundary of the metastable region, in which the system can be found in a local minimum of the thermodynamic potential. The brown and purple curves represent isentropic trajectories in which the system experiences a first-order transition, during which it lives in a mixed phase, until all the volume is occupied by the new stable phase. We also show a few isentropic trajectories passing very close to the CEP.}
	\label{fig:PQMPNJL_PD_zoom1stOrder}
\end{figure*}
%f%f%f%
%f%f%f%
The crossover line ends at a critical endpoint, beyond which the transition becomes of first order, displayed by black solid lines in the figures. The location $(\mu_B^{\rm CEP},T^{\rm CEP})$ of the CEP is found to be (875,121) MeV and (903,118) MeV in the PNJL and PQM model. This can be compared with the results found in independent implementations of the PNJL model~\cite{Contrera:2016rqj}. The corresponding critical isentropes passing through the CEP is given by values of the entropy per baryon $s/n_B=7.02$ and $s/n_B=6.16$, respectively. We decide then to show also a few isentropes in which the system, during its evolution, either passes very close to the CEP or meets the first-order line. This region is shown in finer detail in Fig.~\ref{fig:PQMPNJL_PD_zoom1stOrder}. In this last case, along the critical line, for a given $T$ and $\mu_q$ the thermodynamic potential has two degenerate minima. The single critical line in the $\mu_q\!-\!T$ plane actually becomes the boundary of an extended region of phase coexistence in the $n_q\!-\!T$ plane. In the coexistence region no stable homogeneous phase can exist, but a fraction $\alpha$ of the volume is occupied by the chirally-restored phase and a fraction $(1-\alpha)$ is occupied by the chirally-broken one. They are characterized by the same pressure, temperature and chemical potential, expressing the mechanical, thermal and chemical equilibrium between the two phases.  For each $T$ and $\mu_q$ the value of $\alpha$, which is determined through a Maxwell construction, depends on the history and kind of evolution of the system.
Usually in thermodynamics one considers phase transitions occurring along isothermal lines, however in heavy-ion collisions there is no thermal bath with which the fireball is in contact. The system follows then an isentropic expansion at fixed $s/n_q$ ($n_q=n_B/3$) and one has
\beq
\frac{s}{n_q}=\frac{\alpha s_Q(T,\mu_q)+(1-\alpha)s_H(T,\mu_q)}{\alpha n_Q(T,\mu_q)+(1-\alpha)n_H(T,\mu_q)},
\eeq
where $s_{Q/H}$ and $n_{Q/H}$ are the entropy and quark-number density in the chirally restored/broken phase. One has $\alpha\!=\!1$ when the isentrope crosses the high-density branch of the critical line and $\alpha\!=\!0$ when it crosses the low-density branch. Hence, a given value of $T$ and $\mu_q$ does not fix $\alpha$: one has to specify also the isentrope followed by the system. Notice that, considering this kind of evolution, one is implicitly assuming that the nucleation rate of bubbles of the low-density, chirally-broken phase is larger than the expansion rate of the fireball and we are not addressing the case of super-heating/cooling, which occurs if the system remains in a metastable minimum. Establishing whether this is a realistic assumption of whether the transition occurs via a different dynamics (spinodal decomposition) would need a deeper study of the rate of bubble nucleation, requiring in particular the evaluation of the surface tension of the interface between the two phases~\cite{Randrup:2009gp,Pinto:2012aq,Mintz:2012mz,Fraga:2018cvr}, which is out of the scope of the present paper.
In any case, in Figs.~\ref{fig:PNJL_PD_PDn} and~\ref{fig:PQM_PD_PDn}, we display the regions in which in the two models a homogeneous metastable phase may exist, extending from the continuous first-order line to the dashed spinodal lines.   

The phase diagram plotted in terms of the quark density rather then of the chemical potential reveals how, in both models, the first-order coexistence region extends down to the origin. In the low-temperature regime the behaviour of both models is then unphysical, since they do not leave room for the existence of a self-bound homogeneous nuclear-matter phase, which we know to exist, but whose experimental density $n_B\approx 0.16\,{\rm fm}^{-3}$ would lie here in the coexistence region.
Furthermore there is no room for the liquid-gas phase transition of nuclear matter, which is a characteristic feature of strong interactions in the low-temperature ($T\lsim 20$ MeV), high-density regime for low values of the entropy per baryon~\cite{Bertsch:1983uv,Glendenning:1986zza,Furnstahl:1990zza}.
This must be viewed as a shortcoming of the models due to the pure scalar/pseudoscalar interaction and to the mean field approximation. We expect that the inclusion of a vector interaction and of hadrons as dynamical degrees of freedom in the confined, chirally-broken phase should improve the description of this region of the phase diagram.   

%f%f%f%
\begin{figure*}[!ht]
	\centering
	\includegraphics[width=.49\textwidth]{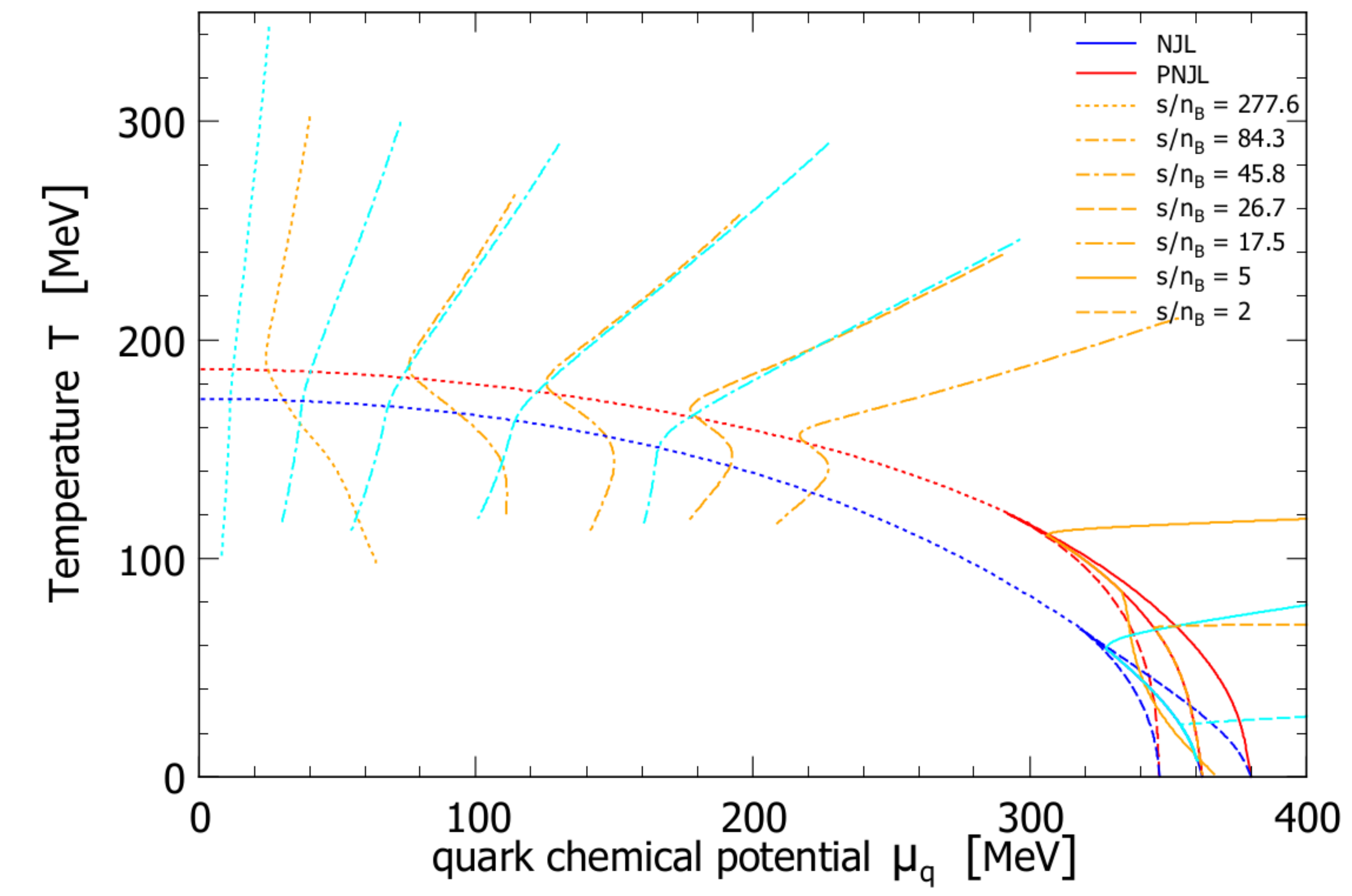}	
	\hfill
	\includegraphics[width=.49\textwidth]{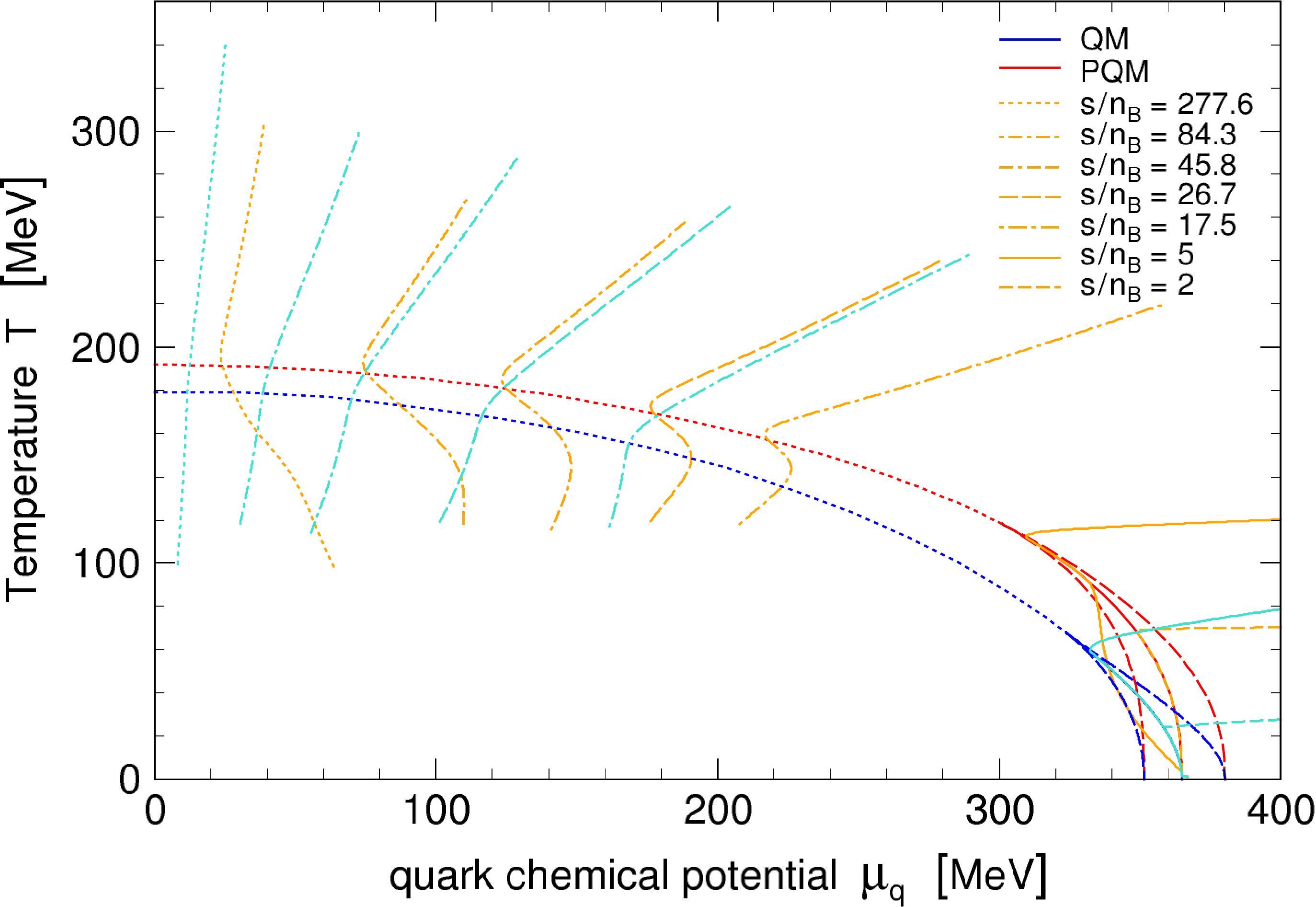}
	\caption{The effect on the phase diagram and on the isentropic trajectories of the Polyakov field. Going from the NJL to the PNJL (left panel) and from the QM to the PQM model (right panel), the CEP moves to higher temperatures and lower chemical potentials. The red/blue curves are the transition lines with/without the coupling between the quarks and the Polyakov field.}
	\label{fig:QMPQM_NJLPNJL_PD}
\end{figure*}
%f%f%f%
In order to assess the role of the Polyakov field in fixing the location of the CEP in the phase diagram, in Fig.~\ref{fig:QMPQM_NJLPNJL_PD} we compare the PNJL vs NJL (left panel) and PQM vs QM models (right panel). In both cases, coupling the quark with the Polyakov field, the CEP turns out to move to a higher temperature and a lower baryo-chemical potential.

\begin{figure*}[!ht]
	\centering
	\includegraphics[width=.49\textwidth]{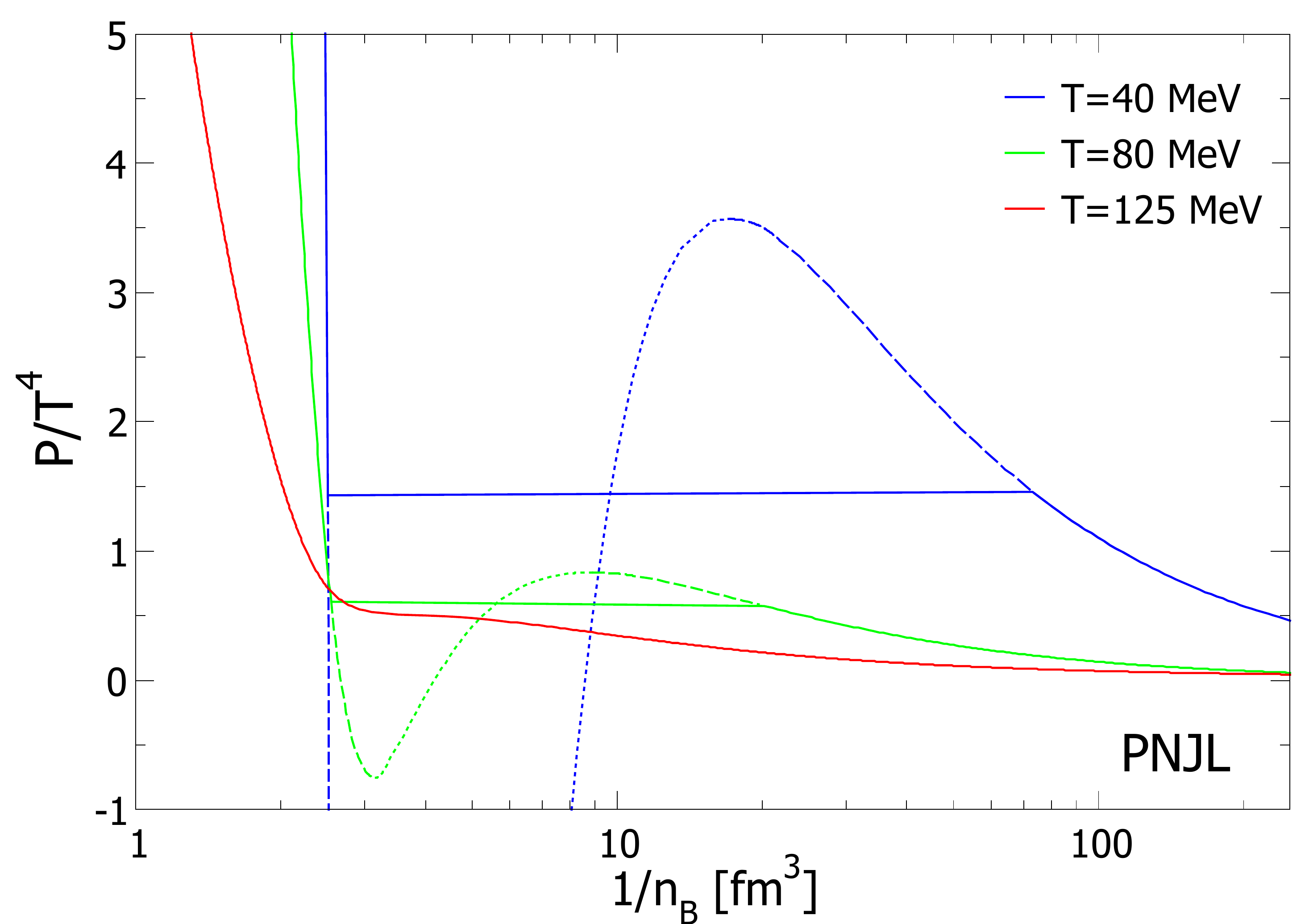}	
	\hfill
	\includegraphics[width=.49\textwidth]{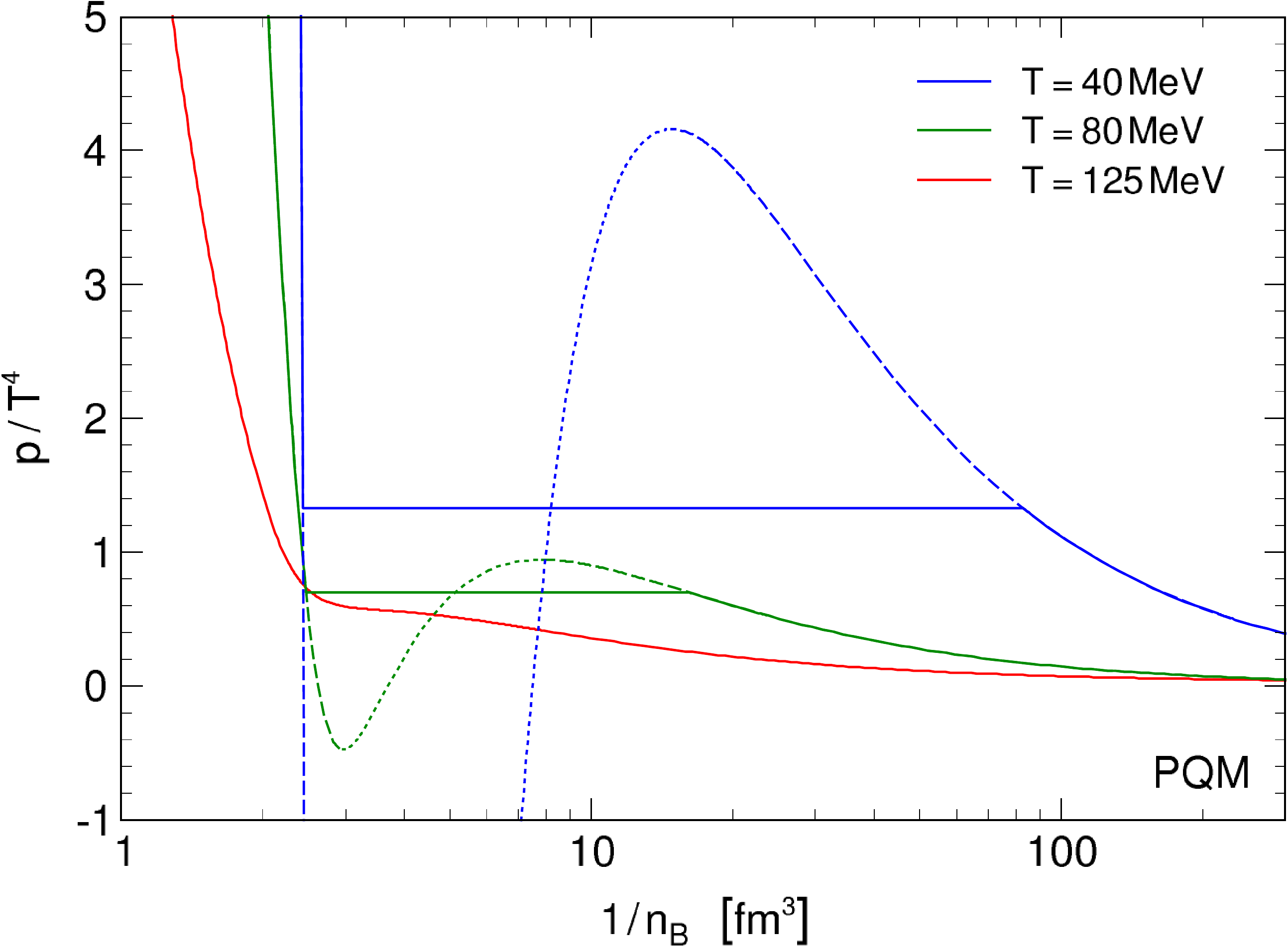}
	\caption{Pressure vs specific volume for a few isothermal transformations in the PNJL (left panel) and PQM (right panel) models. We plot curves both above and below the critical isothermal $T^{\rm CEP}$ passing through the CEP. In the case of a first-order transition we also display the Maxwell construction. The stationary points of the isothermal curves lie on top of the spinodal lines: between them the system is unstable and cannot be found in a homogeneous phase.}
	\label{fig:isothermal}
\end{figure*}
Finally, in Fig.~\ref{fig:isothermal}, we show a few isothermal curves for the two models, plotting the pressure as a function of the specific volume $1/n_B$. We choose values of $T$ either above or below the critical value $T^{\rm CEP}$. In this last case, in which a first-order transition occurs, we show also the Maxwell construction. Notice that the isothermal curves with $T<T^{\rm CEP}$ display two stationary points. Between the two stationary point the pressure is a decreasing function of the density and the system is thus unstable. This part of the curves actually corresponds to effective quark masses arising from the maximum of the thermodynamic potential; this is also the region in which a single homogeneous phase can not exist.
%%%%%%%%%%%%%%%%%%%%%%%%%%%%%%%%%%%%%%%%%%%%%%%%%%%%%%%%%%%%%%%%
\subsection{The speed of sound}
\begin{figure*}[!ht]
	\centering
	\includegraphics[width=.49\textwidth]{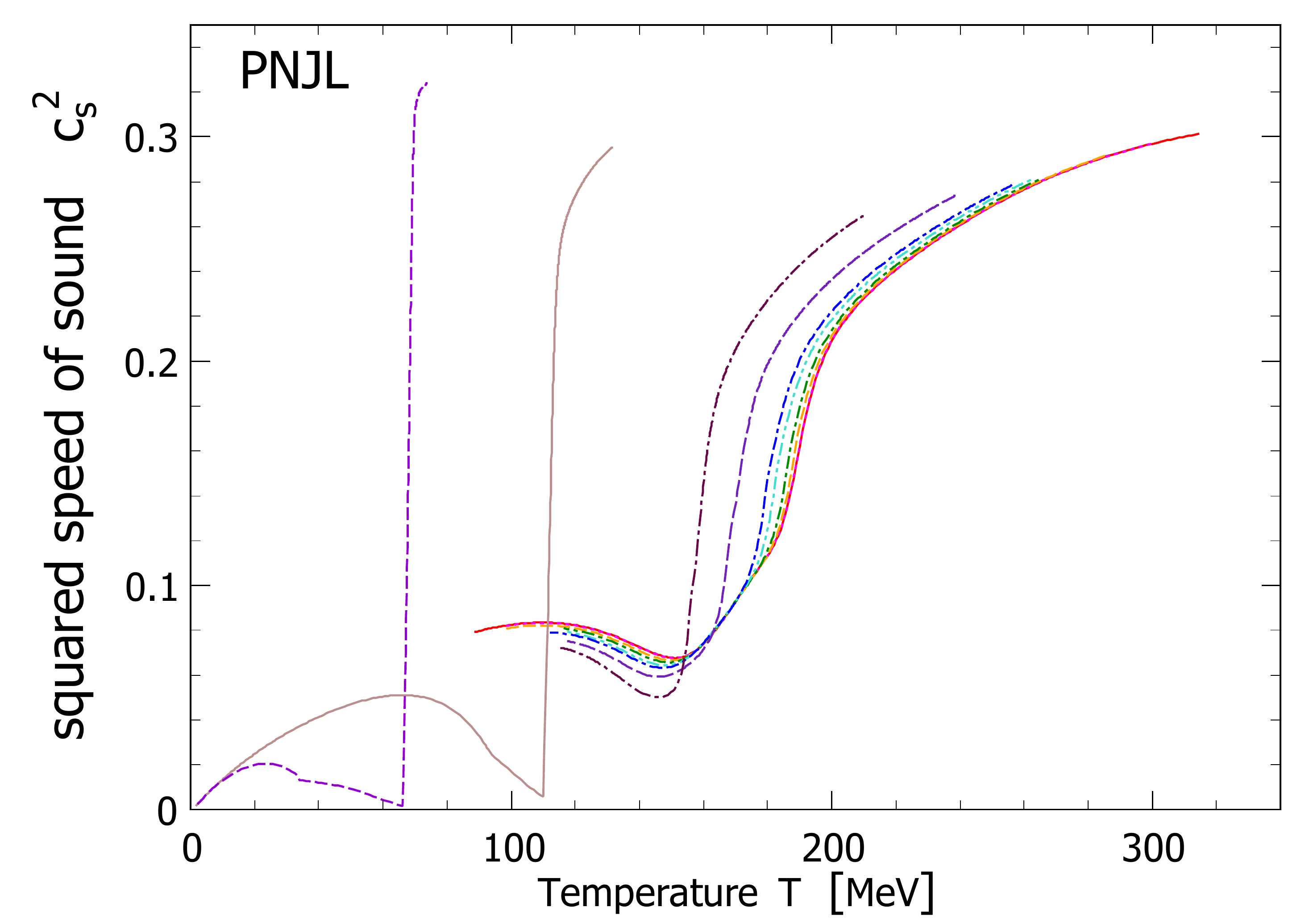}
	\hfill
	\includegraphics[width=.49\textwidth]{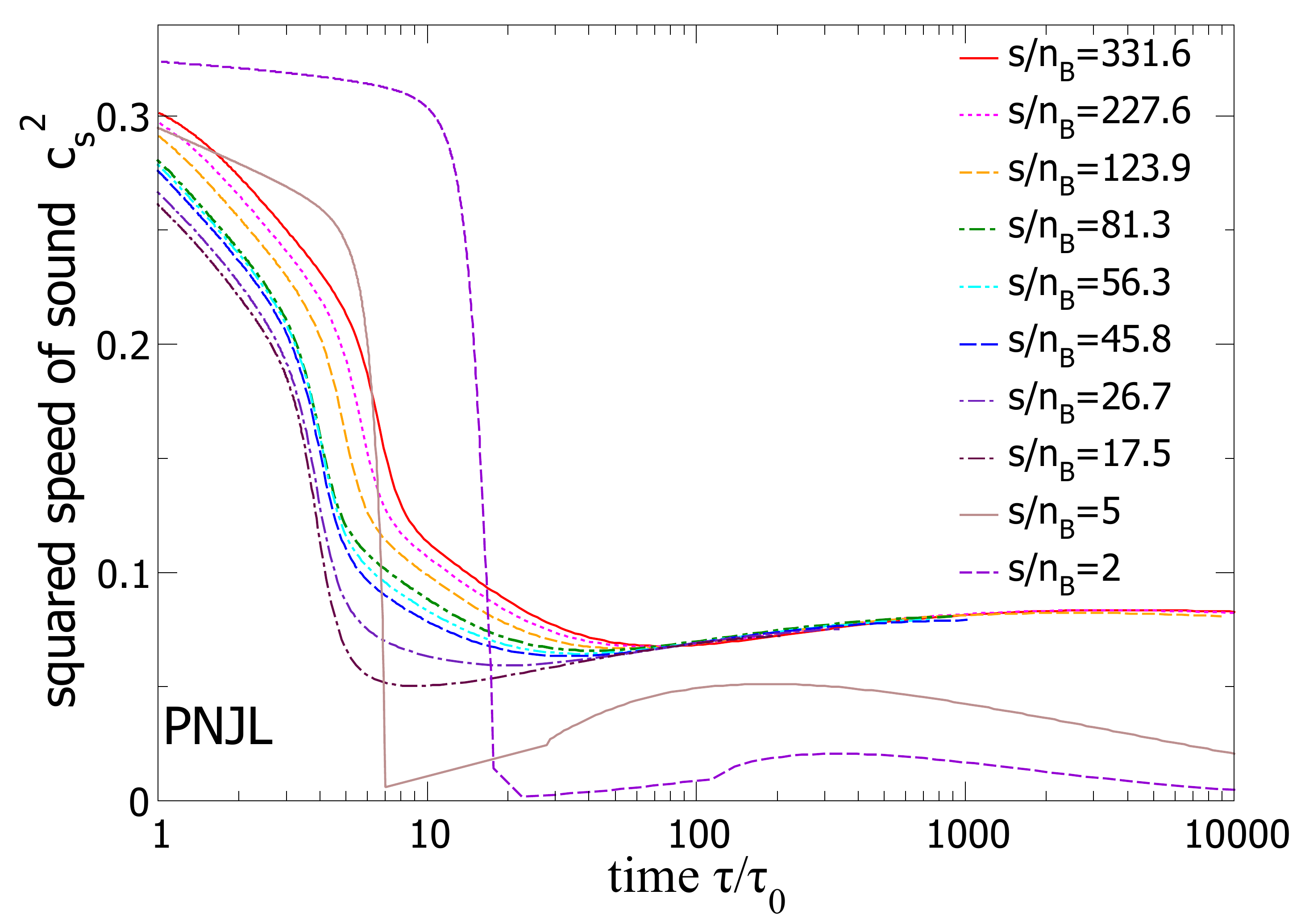}
	\caption[]{The squared speed of sound $c_s^2$ evaluated in the PNJL model along the isentropic trajectories of Fig.~\ref{fig:PNJL_PD_PDn} plotted as a function of temperature (left panel) and time $\tau$ (right panel). The different behaviour between a first-order transition ($s/n_B\!=\!2$, 5) and a smooth crossover is clearly visible.}
	\label{fig:PNJL_cs2}
\end{figure*}
%%%%%%%%%%%%%%%%%%%%%%%%%%%%%%%%%%%%%%%%%%%%%%
\begin{figure*}[!ht]
	\centering
	\includegraphics[width=.49\textwidth]{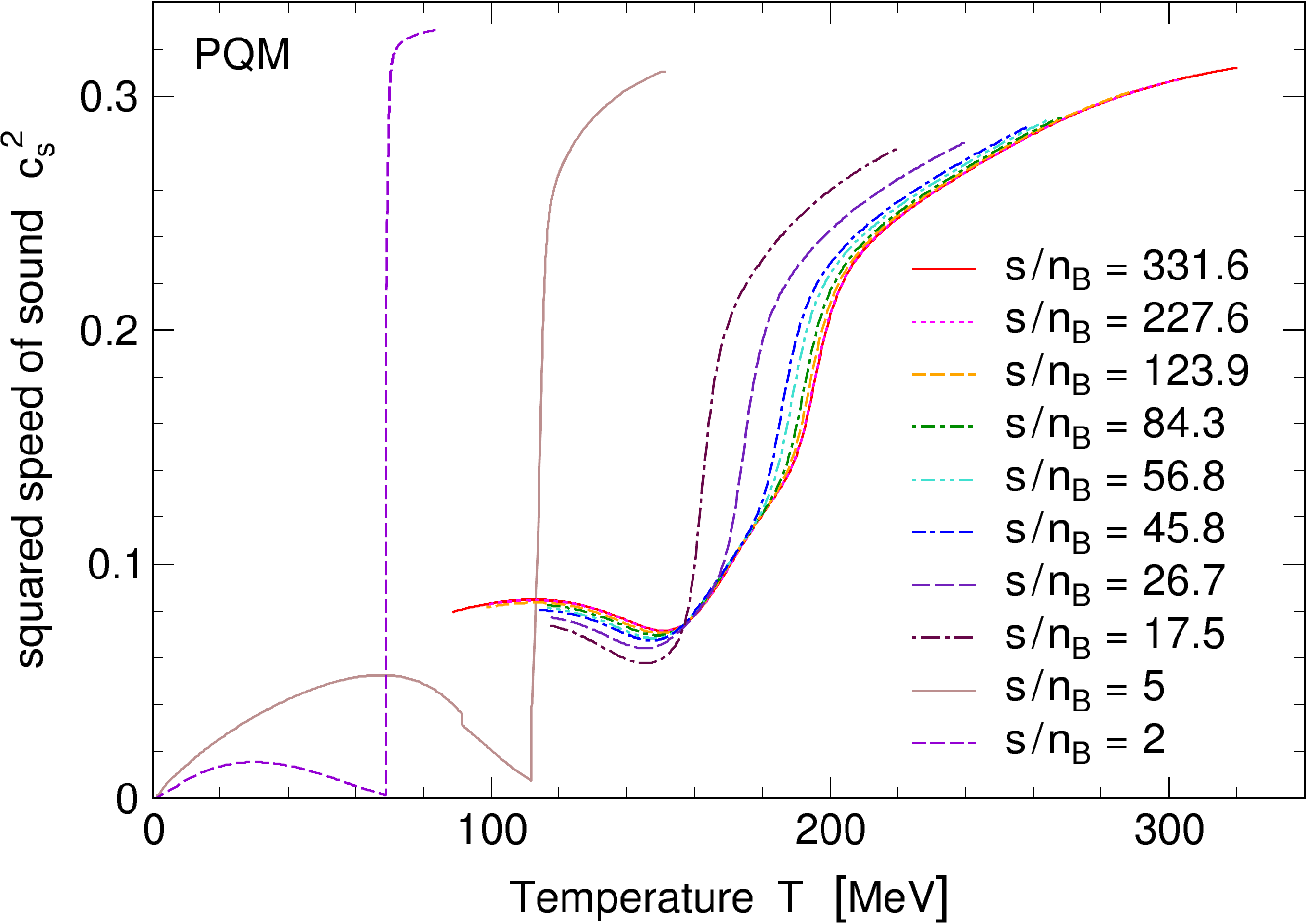}
	\hfill
	\includegraphics[width=.49\textwidth]{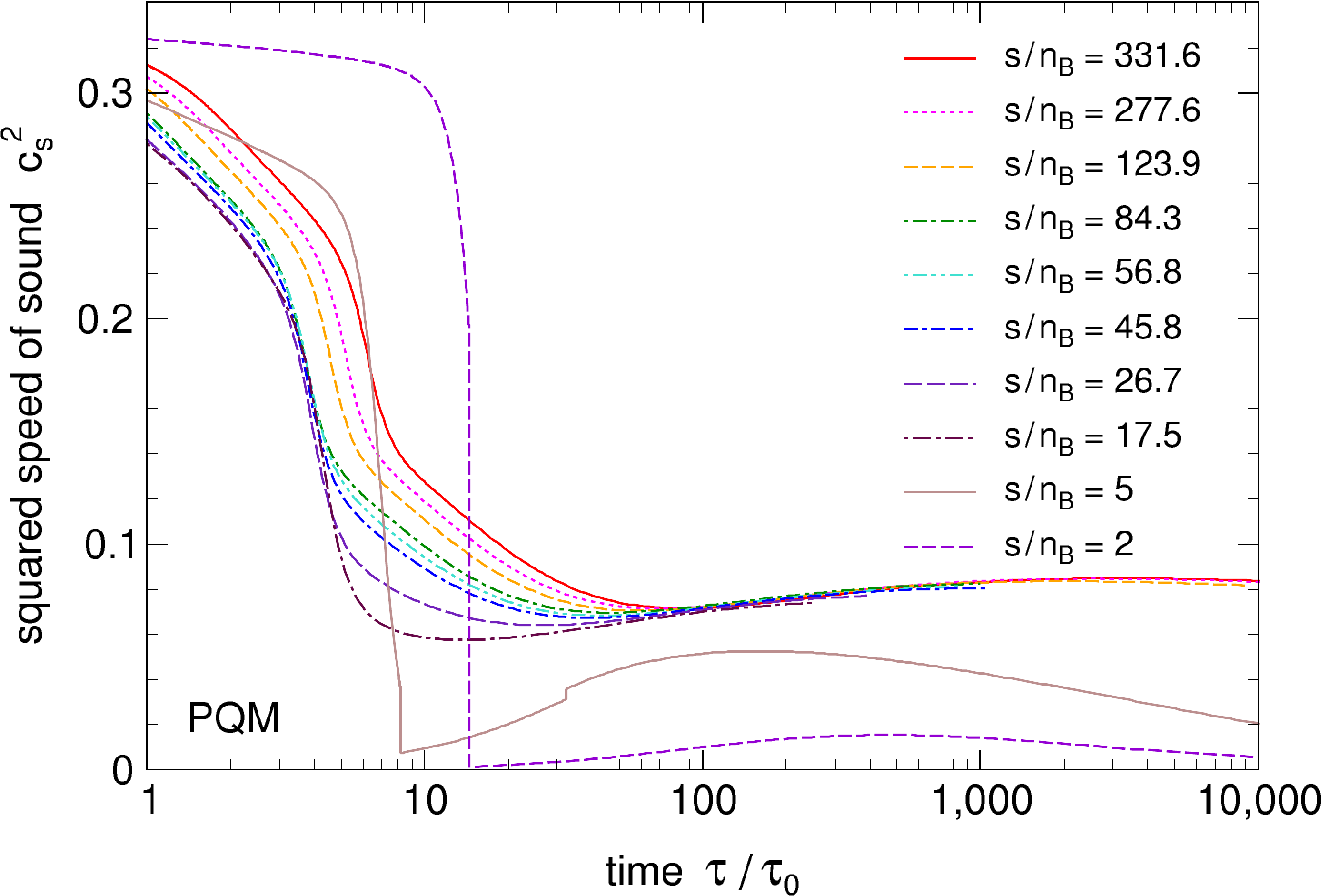}
	\caption[]{The same as in Fig.~\ref{fig:PNJL_cs2}, but for the PQM model.}
	\label{fig:PQMPNJL_cs2}
\end{figure*}
%%%%%%%%%%%%%%%%%%%%%%%%%%%%%%%%%%%%%%%%%%%%%%%%%%%%%%%
When the high temperature of a system allows the
continuous creation/annihilation of particle-antiparticle pairs with $m\ll T$ the mass density in no longer a meaningful concept and the Newtonian definition of speed of sound has to be accordingly generalized.
One can show that the relativistic squared speed-of-sound $c_\mrm{s}^2$ is then given by the following derivative of the pressure with respect to the energy density at constant entropy per particle (baryon in the case of strong interactions):
%e%e%e%
\bea
	c_\mrm{s}^2 &=& \left. \frac{\pd p}{\pd \epsilon} \right|_{s/n}\;.
	\label{eq:cs2}
\eea
The speed of sound maps then a density gradient into a pressure gradient, which -- if the evolution of the system can be described by hydrodynamics -- is responsible for the acceleration of the fluid. 
Having already evaluated the various isentropic trajectories at $s/n_B\!=\!{\rm const}$, the above quantity is simply obtained via a numerical finite-difference method. Results for $c_s^2$ corresponding to the values of $s/n_B$ considered in this work are shown in Figs.~\ref{fig:PNJL_cs2} and~\ref{fig:PQMPNJL_cs2}, plotted as a function of temperature $T$ and time $\tau$. In mapping $T$ into $\tau$ we assume a simple Bjorken-like inviscid longitudinal expansion, for which one has $s\tau=s_0\tau_0$. The curves are plotted down to the kinetic freeze-out temperatures $T_{\rm kin}^{\rm fo}$ provided by the experimental blast-wave fits and quoted in Table~\ref{tab:ise}. Since -- in the mean-field approximation -- as active degrees of freedom in the low-temperature/density phase both models have dressed quarks suppressed by their large effective mass and by their coupling with the Polyakov field, a kinetic freeze-out temperature $T_{\rm kin}^{\rm fo}\!\approx\!120$ MeV corresponds to a very small value of the entropy density, much smaller than the one of a hadron-resonance gas; hence, in order to reach the experimental $T_{\rm kin}^{\rm fo}$, we have to follow the evolution of the system up to unrealistically large values of time $\tau$. Having a more realistic description of the chirally-broken phase within the two models would require to include as dynamical degrees of freedom the set of scalar/pseudoscalar mesons arising from the quark-antiquark interaction contained in the Lagrangian. We leave this issue for future work.

In Figs.~\ref{fig:PNJL_cs2} and~\ref{fig:PQMPNJL_cs2} one can appreciate the very different behaviour of the speed of sound depending whether, during the isentropic evolution, the transition from the chirally restored to the broken phase occurs via a smooth crossover (curves with $s/n_B\ge 17.5$) or via a first-order change of state (curves with $s/n_B=2$, 5). For the $s/n_B$ values of experimental relevance in both models the transition occurs in the crossover region and the speed of sound displays a rapid but continuous drop for temperatures around the inflection point of the light chiral condensate. Notice that there is no qualitative difference in the behaviour of the curves corresponding to the various center-of-mass energies explored at RHIC, although -- at a quantitative level -- for all values of time the speed of sound for low $s/n_B$ is significantly lower than the one for large $s/n_B$: this should have an impact on the final momentum distributions of soft hadrons. 

On the other hand if, during its evolution, the system meets a first-order transition the speed of sound suddenly drops (almost) to zero. Such a soft equation of state (EoS) should translate into a very small acceleration of the fluid during this stage, leaving its signatures in the final momentum distributions of the hadrons decoupling from the fireball.

More quantitative considerations would require inserting the above EoS into a full hydrodynamic code including also the evolution of baryon density~\cite{Du:2019obx}, allowing one to simulate the expansion of the fireball and the continuous decoupling of light hadrons.
%f%f%f%

\subsection{Generalized susceptibilities}
%f%f%f%
\begin{figure*}[!ht]
	\centering
	\includegraphics[width=.49\textwidth]{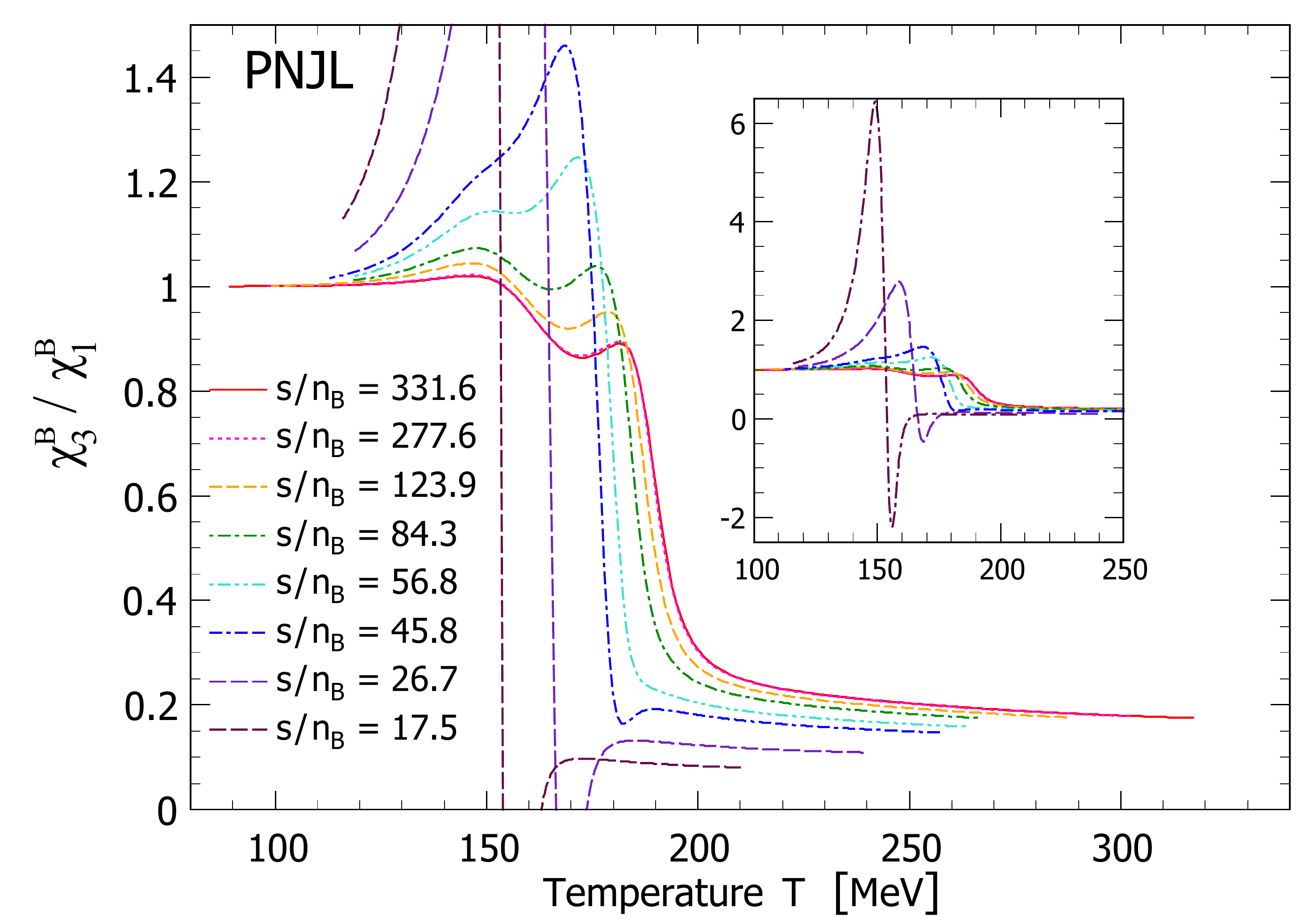}
	\hfill
	\includegraphics[width=.49\textwidth]{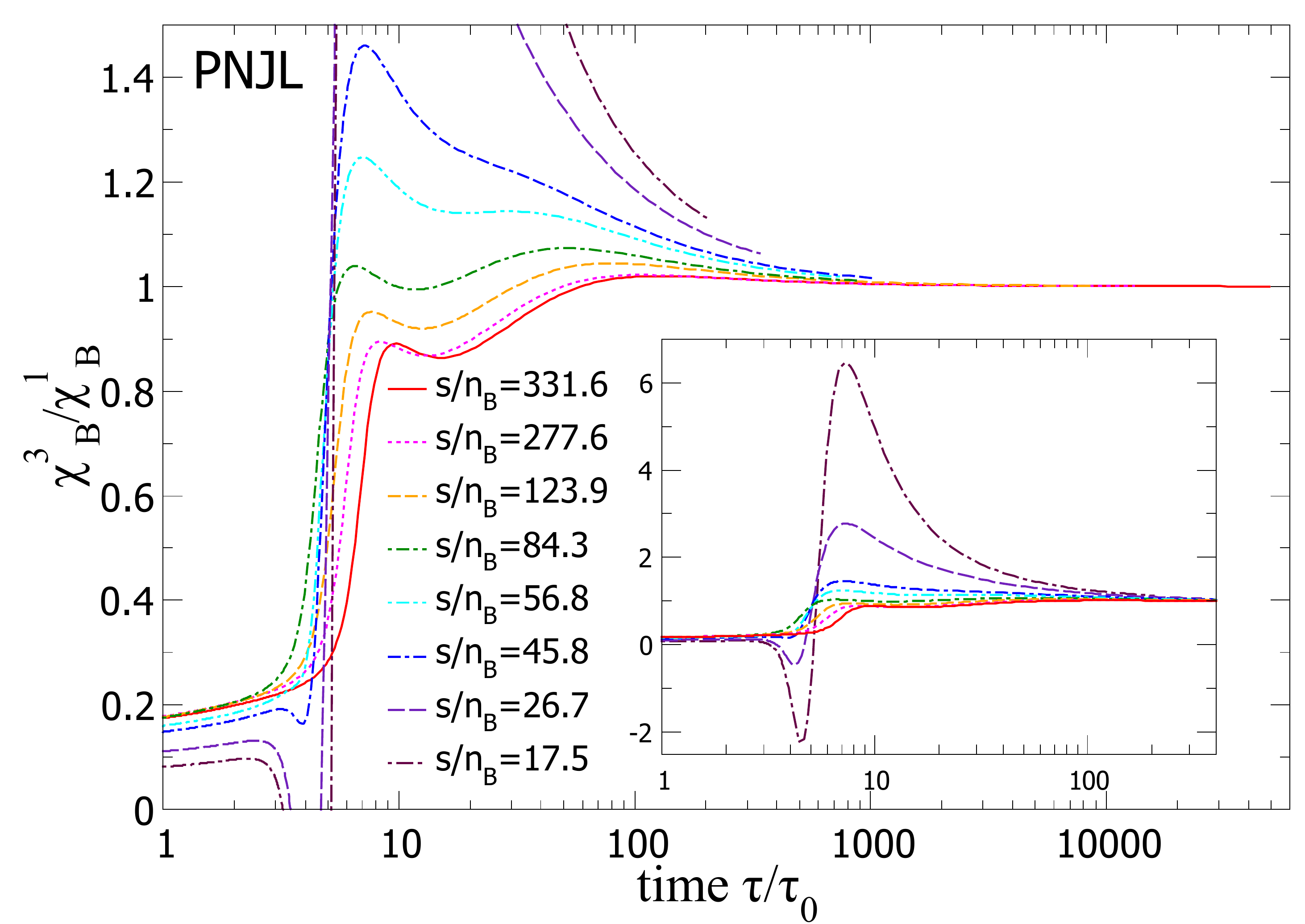}
	\caption[]{PNJL results for the normalized skewness of the baryon-number distribution plotted as a function of temperature (left panel) and time $\tau$ (right panel) along the isentropic trajectories shown in \Fig{fig:PNJL_PD_PDn}. The evolution always starts at time $\tau_0=1$ fm/c, corresponding to the entropy densities $s_0$ given in Table~\ref{tab:ise}.}
	\label{fig:PNJL_skewness}
\end{figure*}
%f%f%f%
%f%f%f%
\begin{figure*}[!ht]
	\centering
	\includegraphics[width=.49\textwidth]{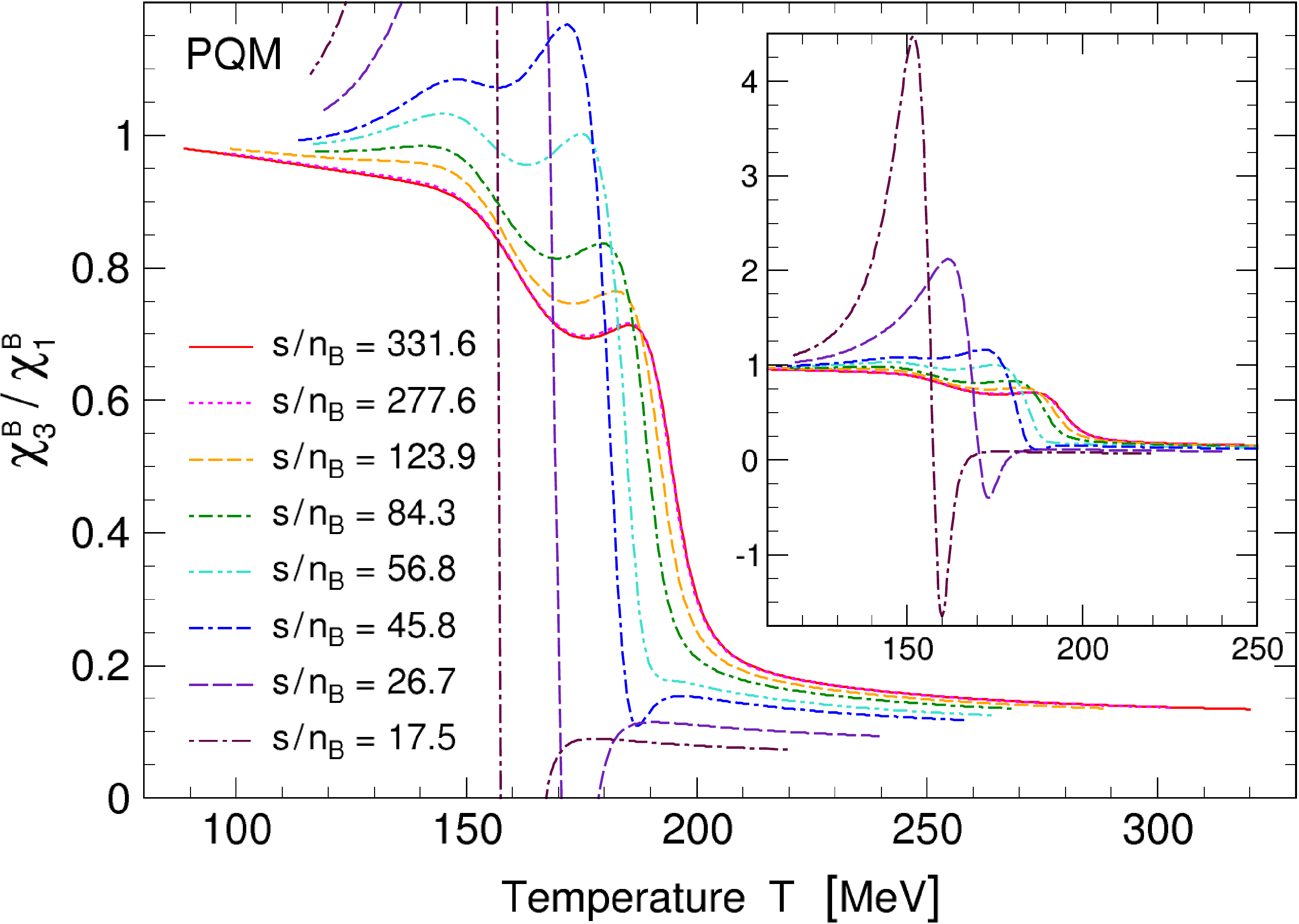}
	\hfill
	\includegraphics[width=.49\textwidth]{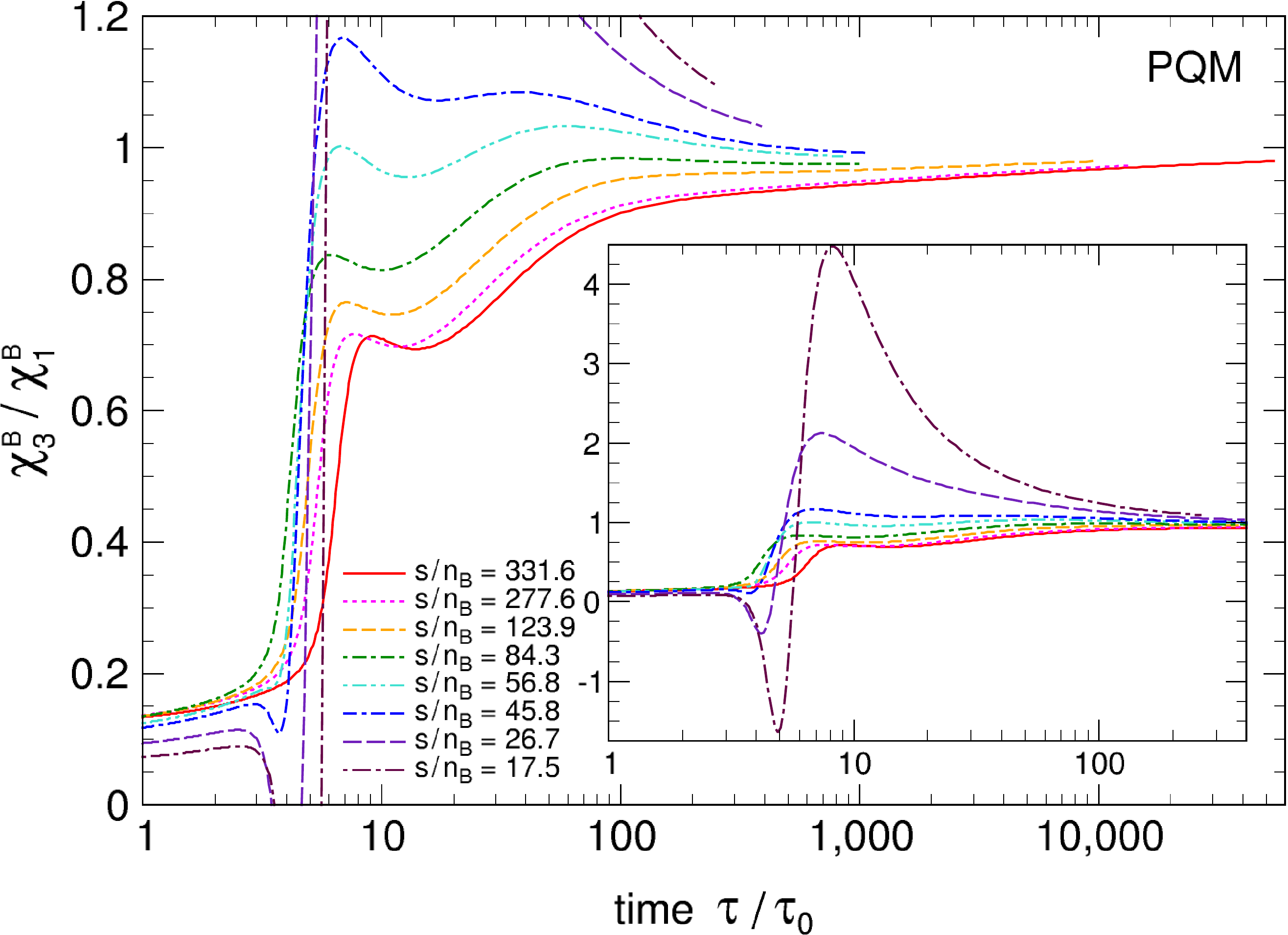}%\\
	\caption[]{The same as in Fig.~\ref{fig:PNJL_skewness}, but for the PQM model.}
	\label{fig:PQMPNJL_skewness}
\end{figure*}
%f%f%f%

%f%f%f%
\begin{figure*}[!ht]
	\centering
	\includegraphics[width=.49\textwidth]{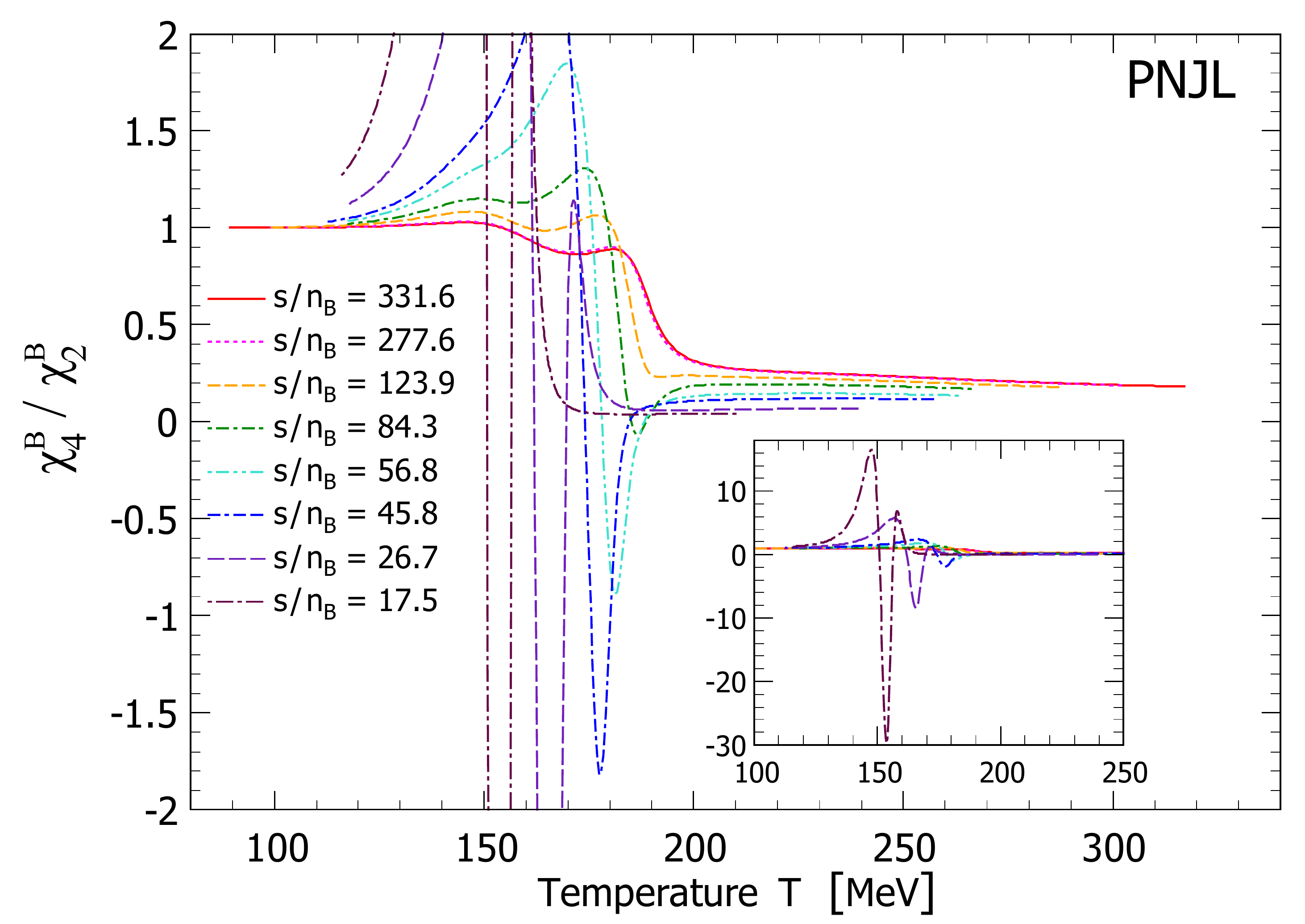}
	\hfill
	\includegraphics[width=.49\textwidth]{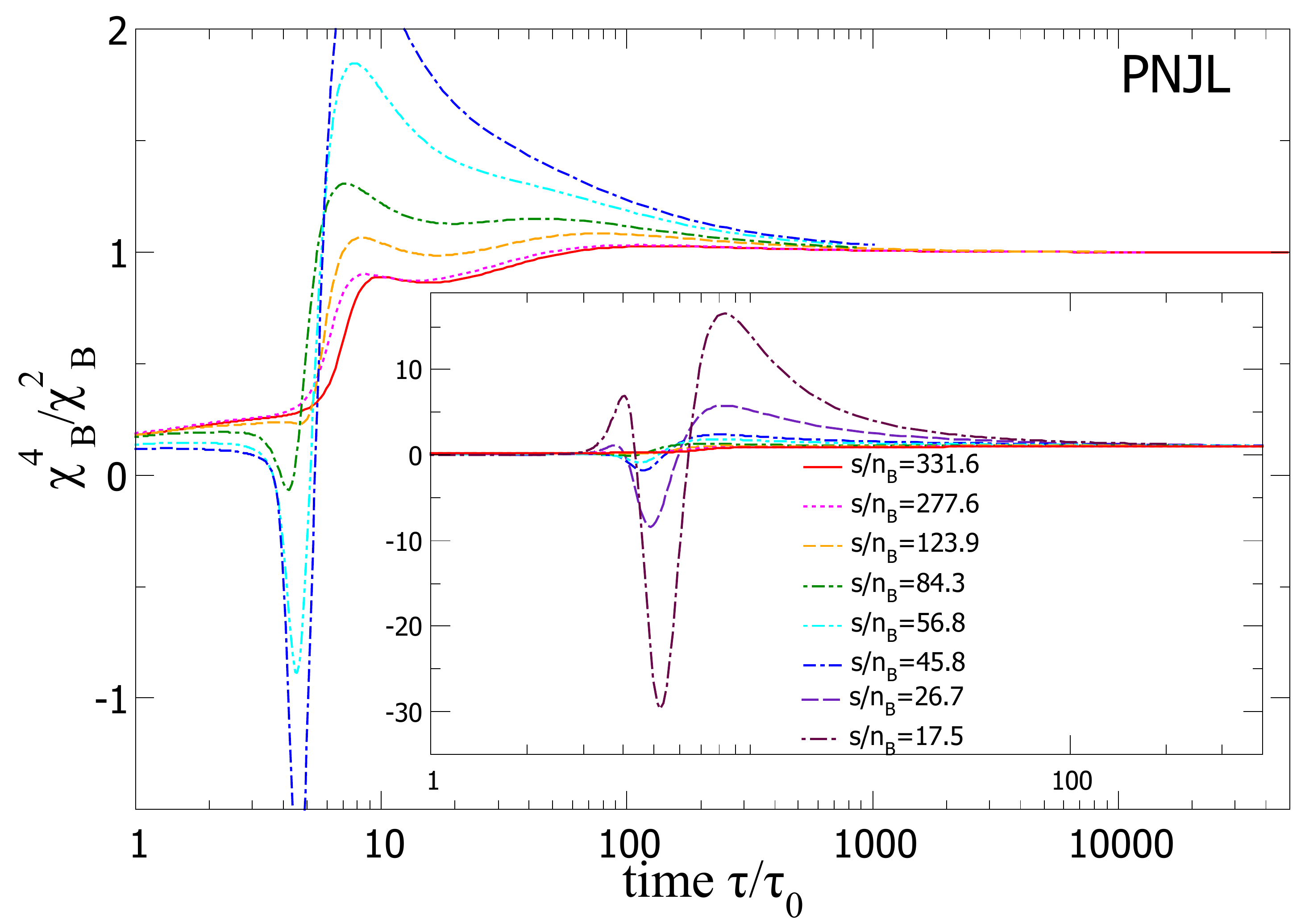}
	\caption[]{PNJL results for the normalized kurtosis of the baryon-number distribution plotted as a function of temperature (left panel) and time $\tau$ (right panel) along the isentropic trajectories shown in \Fig{fig:PNJL_PD_PDn}. The evolution always starts at time $\tau_0=1$ fm/c, corresponding to the entropy densities $s_0$ given in Table~\ref{tab:ise}.}
	\label{fig:PNJL_kurtosis}
\end{figure*}
%f%f%f%
%f%f%f%
\begin{figure*}[!ht]
	\centering
	\includegraphics[width=.49\textwidth]{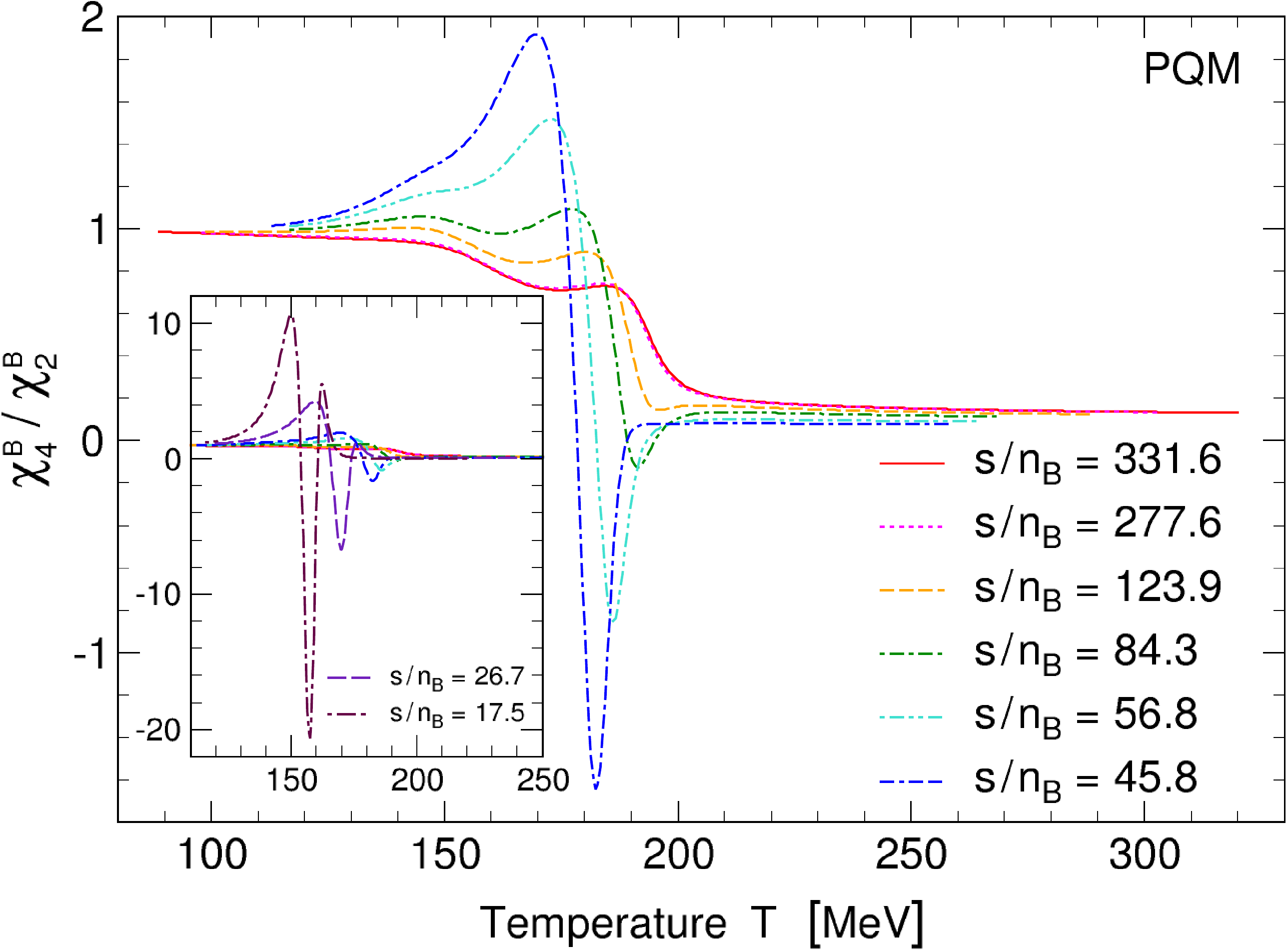}
	\hfill
	\includegraphics[width=.49\textwidth]{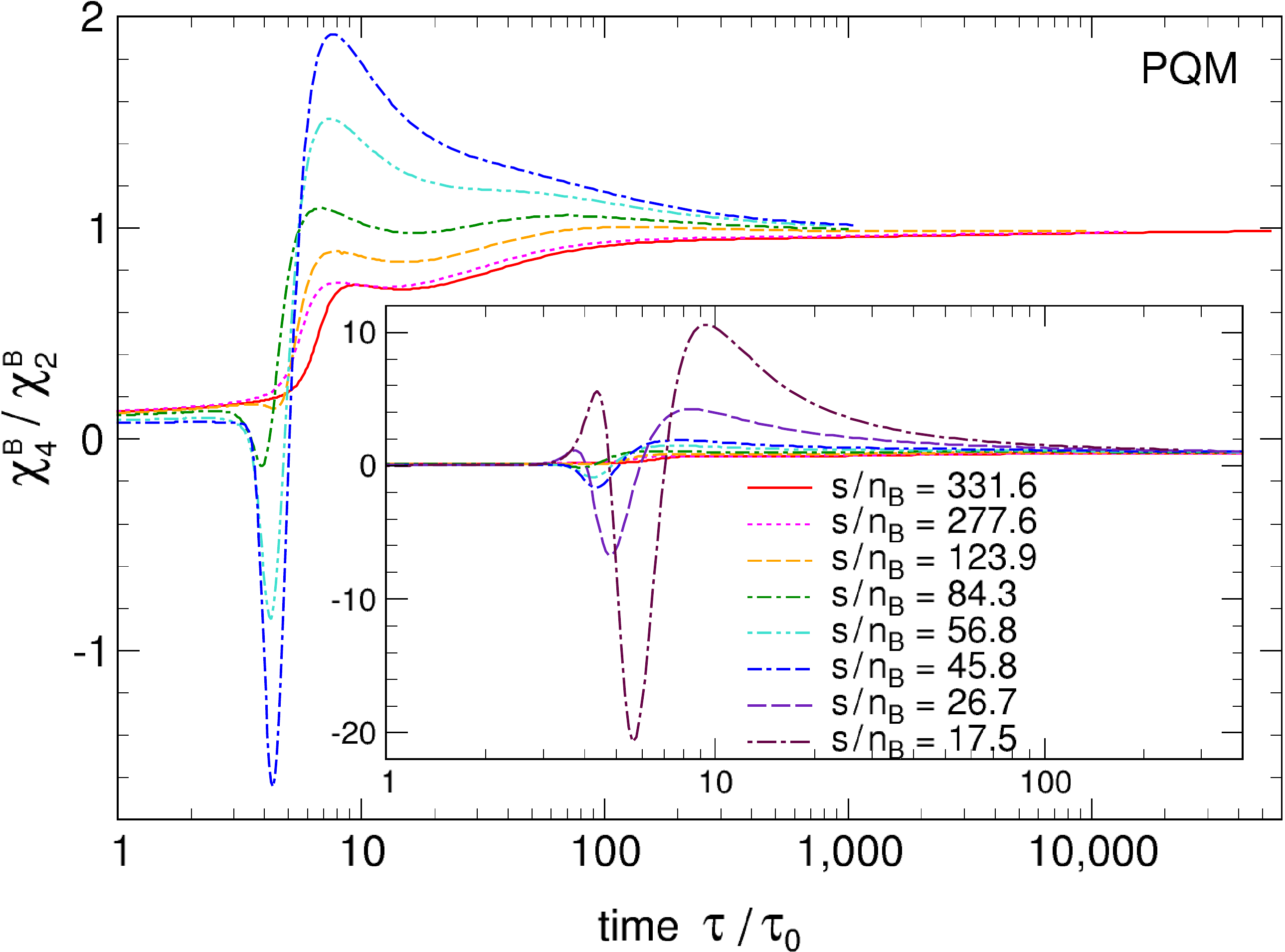}%\\
	\caption[]{The same as in Fig.~\ref{fig:PNJL_kurtosis}, but for the PQM model.}
	\label{fig:PQMPNJL_kurtosis}
\end{figure*}
%f%f%f%
In this section we evaluate higher-order cumulants of the distributions of conserved charges (in particular, baryon number) provided by the PNJL and the PQM models. For the sake of consistency, here we remind the basic definitions. In statistical mechanics cumulants are expressed in terms of derivatives of the pressure with respect to the chemical potential, i.e~\cite{Asakawa:2015ybt}.
\beq
\langle \hat N^n\rangle_c\equiv\left.\frac{\partial^n(-\Omega/T)}{\partial(\mu/T)^n}\right|_T=\left.\frac{\partial^n(P/T)}{\partial(\mu/T)^n}\right|_TV\label{eq:cum-def},
\eeq
where $\hat N$ is the conserved charge and $\mu$ the associated chemical potential. In heavy-ion collisions the relevant conserved charges are baryon number, strangeness and electric charge, however here we simply focus on baryon-number fluctuations. The first few cumulants of a distribution are given by ($\delta\hat N\equiv\hat N-\langle\hat N\rangle$)
\begin{multline}
\langle\hat N\rangle_c=\langle\hat N\rangle,\quad
\langle\hat N^2\rangle_c=\langle\delta\hat N^2\rangle,\\
\langle\hat N^3\rangle_c=\langle\delta\hat N^3\rangle,\quad
\langle\hat N^4\rangle_c=\langle\delta\hat N^4\rangle-3(\langle\delta\hat N^2\rangle)^2
\end{multline}
and allow one to define its mean $M$, its variance $\sigma^2$, its skewness $S$ and kurtosis $\kappa$
\beq
M\equiv\langle\hat N\rangle_c,\;\sigma^2\equiv\langle\hat N^2\rangle_c,\;  S\equiv\frac{\langle\hat N^3\rangle_c}{\langle\hat N^2\rangle_c^{3/2}}\;{\rm and}\;
\kappa\equiv\frac{\langle\hat N^4\rangle_c}{\langle\hat N^2\rangle_c^{2}}.
\eeq
In particular, skewness and kurtosis of a distribution quantify how asymmetric and peaked/broad the latter is with respect to a Gaussian, for which $\langle \hat N^n\rangle_c=0$ for $n\ge 3$ and hence $S=\kappa=0$.

Why is the study of higher order cumulants of interest for heavy-ion collisions?
A first motivation comes from the search of the CEP in the QCD phase-diagram, where -- if present -- the transition is of second order and hence characterized by an infinite correlation length of the order parameter, in this case the chiral condensate $(\bar\psi\psi)$. One finds that higher order cumulants display a stronger sensitivity to such a quantity~\cite{Stephanov:2008qz} and this is of relevance for experimental studies, in which the finite size and the expansion rate of the produced medium -- far from the thermodynamic limit in which a phase transition is rigorously defined -- would prevent one from observing any actual divergence of a correlation length.
Secondly, the experimental measurement of higher-order fluctuations of conserved charges in heavy-ion collisions and the comparison with theory results provided by lattice-QCD simulations and Hadron-Resonance Gas model calculations allow one to get an independent estimate of the chemical freeze-out point~\cite{Karsch:2012wm,Borsanyi:2014ewa,Alba:2014eba}, besides the usual one based on the average yields of the various hadronic species.
Finally, various combinations of cumulants of conserved charges, allow one to extract information on the nature of charge carriers, and hence of the active degrees of freedom (hadrons or quarks in the case of QCD), in the medium: for a comprehensive review see Ref.~\cite{Asakawa:2015ybt}.

Being the cumulants extensive quantities it is often convenient to take ratios of the latter, so to cancel the dependence on the volume $V$ in Eq.~(\ref{eq:cum-def}), which can be poorly known: in particular, in the case of relativistic heavy-ion collisions, the only indirect information on the size of the medium comes from the multiplicity of produced particles and from HBT correlations. Furthermore one prefers dealing with dimensionless quantities, in which the trivial $T^4$ dependence of the pressure of a relativistic plasma is factorized. Ratios of cumulants (here, simply of the baryon-number distribution) are then defined in terms of ratios of the following generalized susceptibilities:
\begin{equation}
    \chi^B_n\equiv\left.\frac{\partial^n (P/T^4)}{\partial (\mu_B/T)^n}\right|_T.
\end{equation}
Here, as independently done in the literature (see e.g.~\cite{Ferreira:2018sun}), we focus on the following two interesting ratios of cumulants, directly related to the skewness and the kurtosis of the baryon-number distribution:
\beq
\frac{\chi^B_3}{\chi^B_1}=S\frac{\sigma^3}{M}\quad{\rm and}
\quad\frac{\chi^B_4}{\chi^B_2}=\kappa\sigma^2.
\eeq
For the sake of simplicity in the following, with abuse of language, we often refer to the above ratios as normalized skewness and kurtosis.
Their interest is first of all related to their behaviour around the chiral transition. The second order cumulant $\chi_2^B$ displays a ridge structure in the $\mu_B-T$ plane along the crossover line~\cite{Asakawa:2015ybt}. Hence, we expect that $\chi_3^B$ changes sign and $\chi_4^B$ stays negative around the location of the chiral crossover. This is what one actually observes in Figs.~\ref{fig:PNJL_skewness}-\ref{fig:PQMPNJL_kurtosis}, where $\chi_3^B/\chi_1^B$ and $\chi_4^B/\chi_2^B$ are plotted along the isentropic trajectories corresponding to values of $s/n_B$ of interest for the ongoing BES at RHIC. There is in fact a rich experimental activity aiming at finding a  non-monotonic behaviour of the skewness and kurtosis of net-proton, kaon and charge distributions as one varies the center-of-masss energy of the collision~\cite{Adamczyk:2013dal,Adamczyk:2014fia}, although no definite conclusions can be drawn so far. 

As already mentioned, ratios of higher-order cumulants allows one to get information on the nature of the active degrees of freedom in the medium under the different conditions of temperature and density.
Consider first the classical limit, in which for the (quasi-)particles of the medium the condition $E_p-\mu\gg T$ holds. In this case 
\beq
P\sim\frac{m^2T^2}{2\pi^2}e^{\mu/T}K_2\left(\frac{m}{T}\right)
\eeq
From Eq.~(\ref{eq:cum-def}) it follows that particles obey a Poissonian distributions, with all cumulants equal to the mean, $\langle N^n\rangle_c=M$. Considering the fluctuations of conserved charges in a relativistic gas one is actually interested in the distribution of the net number of particles (particles minus antiparticles). One gets then the difference of two Poissonian, i.e. a Skellam distribution, for which one has:
\beq
\langle N_{\rm net}^n\rangle_c=\langle N\rangle+(-1)^n\langle\overline N\rangle,
\eeq
where the factor $(-1)^n$ comes from the multiple derivatives with respect to the chemical potential of the antiparticles, entering with a minus sign. Consider now the following ratio of higher-order cumulants of the net baryon number:
\beq
\frac{\langle N_{B,{\rm net}}^{n+2}\rangle_c}{\langle N_{B,{\rm net}}^n\rangle_c}=
\frac{B^{n+2}[\langle N_B\rangle+(-1)^{n}\langle\overline N_B\rangle]}{B^{n}[\langle N_B\rangle+(-1)^n\langle\overline N_B\rangle]}=B^2\label{eq:dof},
\eeq
where $B$ is the baryon charge carried by the elementary active degrees of freedom. We expect this ratio to be 1 in the confined, chirally-broken phase, where baryon number is carried by particles -- the baryons -- with $B=1$.
In Figs.~\ref{fig:PNJL_skewness}-\ref{fig:PQMPNJL_kurtosis} we show that in the low-temperature/density phase $\chi_3^B/\chi_1^B$ and $\chi_4^B/\chi_2^2$ actually approach 1 both in the PNJL and PQM models. From Eqs.~(\ref{eq:effective-q}) and~(\ref{eq:effective-qbar}) one can see that this arises from the role of the Polyakov field, which implements in an effective way colour-confinement suppressing the contribution to the pressure from single-quark ($B=1/3$) and two-quark ($B=2/3$) clusters, leaving unquenched only the one from states with three quarks of different colours and total baryon number $B=1$.
We stress that, if the effective valence quarks, with $B=1/3$, were active degrees of freedom one would get a much lower value for this ratio, close to $1/9$: this is what actually happens in the standard NJL and QM models, which do not implement quark confinement.
Notice however that, both in the PNJL and PQM models, in the high-temperature/density limit in which baryon number is carried by quarks, $\chi_3^B/\chi_1^B$ and $\chi_4^B/\chi_2^2$ saturate to a very low value, even smaller than the classical result 1/9 given by Eq.~(\ref{eq:dof}), due to the effect of quantum statistics.

%%%%%%%%%%%%%%%%%%%%%%%%%%%%%%%%%%%%%%%%%%%%%%%%%%%%%%%%%%%%%%%%%
\section{Discussion and perspectives}\label{sec:conclusions}
In this paper we explored the thermodynamics of strongly-interacting matter through two effective Lagrangians developed in the literature to describe the spontaneous breaking of $Z_3$ (confinement/deconfinement transition) and chiral symmetry: the PNJL and the PQM models. We performed our study in the mean-field approximation, describing the system as a gas of quarks endowed with effective masses and coupled to the Polyakov fields. Both the effective quark masses and the expectation value of the Polyakov fields are obtained requesting the thermodynamic potential to be stationary under variations of the above quantities.
Our aim was to obtain very general qualitative information on the phase-diagram of strong interactions and on the behaviour of matter in the different regions of the $\mu_B\!-\!T$ plane, not accessible by lattice-QCD simulations, limited to the case of vanishing or very small baryon density. Our phenomenological motivation was to provide a theoretical guidance for the rich ongoing and future experimental programs at RHIC, SPS, NICA and FAIR, in which heavy-ion collisions at low center-of-mass energy (will) allow the exploration of the chiral/deconfiment transition in the high-density region of the QCD phase-diagram.

The strongest motivation of the above experimental program is the search for the possible critical endpoint in the QCD phase-diagram, were the sharper and sharper crossover would turn into a first-order transition. Hence, we started our analysis identifying the crossover/first-order transition line in the $\mu_B\!-\!T$ plane and establishing the location of the CEP: for  $(\mu_{B}^\mrm{CEP},T^\mrm{CEP})$ we obtained the values $(875,121)$ MeV and $(903,118)$ MeV in the PNJL and PQM models, respectively. It is of interest to check how close/far these values are from the region currently explored in Beam-Energy Scan ongoing at RHIC. QCD conserves baryon number; furthermore, if dissipative effects are small, entropy production is negligible during the expansion of the fireball arising from the heavy-ion collisions. Hence, the evolution of the medium produced in the collision occurs along trajectories of constant entropy-per-baryon. We estimated the $S/BB$ values of relevance for the BES at RHIC starting from the measured yields of charged particles (assuming $S\sim N^{\rm ch}$) and net protons, obtaining values ranging from 17 to 330 at the lowest ($\sqrt{s_{\rm NN}}\!=\!7.7$ GeV) and highest ($\sqrt{s_{\rm NN}}\!=\!200$ GeV) center-of-mass energy respectively. For comparison, in the PNJL and PQM models, the critical isentropes passing through the CEP correspond to values of $(s/n_B)_{\rm crit}$ of 7.02 and 6.16. If the above estimates are realistic one should conclude that at RHIC we have not yet reached conditions close to the CEP of the QCD phase-diagram. 
Due to its conceptual and possibly phenomenological importance, in our analysis we explored also the first-order transition region, where two different phases can coexist in equilibrium or -- if the evolution of the system is rapid enough compared to the bubble nucleation rate -- the system can remain for long in a metastable phase. Hence, in our figures, we also plotted the corresponding spinodal curves predicted by the two models, boundaries of the metastability regions. 

If, even at the lowest center-of-mass energy explored in the BES at RHIC, the transition might occur far from the first-order region, are there experimental signatures sensitive to the steeper and steeper crossover as the baryon density increases? We started our investigation considering the speed of sound, whose evolution was studied along the different isentropes. The qualitative behaviour of all the curves referring to the crossover case looks very similar, however -- at the quantitative level -- at any given time the value of $c_s^2$ in collisions at lower center-of-mass energies is always much lower than the one at higher $\sqrt{s_{\rm NN}}$. Since $c_s^2$ maps energy-density gradients into pressure gradients, responsible for the fluid acceleration, such a sizeable softening of the EoS at lower values of $\sqrt{s_{\rm NN}}$ should strongly affect the flow of the medium.
Experimental signatures of the softening of the EoS can be looked for in the transverse-momentum distributions of the produced hadrons. The average transverse mass $\langle m_\perp\rangle$ of pions, kaons and protons as a function of $\sqrt{s_{\rm NN}}$ was studied in nucleus-nucleus collisions at AGS, SPS, RHIC and LHC. One found $\langle m_\perp\rangle-m\approx 0.2$ GeV at the low AGS energies~\cite{Ahle:1999uy}, a flat plateau around a higher value at SPS~\cite{Alt:2007aa} and at the lowest energies of the BES at RHIC~\cite{Adamczyk:2017iwn} and an increasing trend starting from $\sqrt{s_{\rm NN}}\approx 60$ GeV up to LHC energies. These findings look compatible with the combined effect of the softening of the EoS and of the milder energy-density gradients as $\sqrt{s_{\rm NN}}$ gets lower. Actually, people suggested that the flat plateau of $\langle m_\perp\rangle$ at SPS energies might be due to the coexistence of the hadronic and deconfined phases typical of a first-order transition.
However, more solid conclusions can only be drawn solving the full set of hydrodynamic equations for finite baryon density and with a realistic EoS, with a steeper and steeper crossover eventually turning into a first-order transition.

If the speed of sound mainly affects the momentum distribution of the produced hadrons, the fluctuations of their yields -- more precisely the ones associated to conserved quantities, like net protons, kaons and charged particles -- are related to the higher-order susceptibilities of baryon-number, strangeness and electric charge. Accordingly, we addressed their evaluation starting from the mean-field thermodynamic potential of the PNJL and PQM models, focusing in this paper only on baryon-number fluctuations. At the CEP the order parameter (the $\sigma$ field or chiral condensate) is characterized by an infinite correlation length $\xi$ leading to a divergence of its cumulants and of all the cumulants of quantities coupled to the latter, like the baryon number. Notice that in heavy-ion collisions, due to the finite size and lifetime of the medium, one would not observe any actual divergence of the correlation length and it is thus necessary to focus on higher-order cumulants, which display a stronger sensitivity on $\xi$. Hence, we studied the ratios of the generalized baryon-number susceptibilities $\chi_3^B/\chi_1^B$ and $\chi_4^B/\chi_2^B$ along lines of fixed $s/n_B$ of interest for the BES at RHIC, finding that for the trajectories passing closer to the CEP -- for which the crossover is steeper -- the above quantities display sizeable oscillations and rapid changes of sign in the transition region. 
Clearly, one aims at finding signatures of the above non-trivial behaviour in the proximity of the CEP in the experimental fluctuations of conserved charges (baryon number, strangeness and electric charge), looking for deviations from a trivial Skellam statistics, although how far the chemical freeze-out point lies from the chiral transition plays a crucial role. For the moment the above higher-order cumulants were used for a less ambitious purpose, namely to get an independent estimate of the chemical freeze-out point in the $\mu_B-T$ plane, besides the usual one based on the average yields of identified hadrons.

Our work has to be considered just as a first step towards the aim of providing solid theoretical qualitative guidance within chiral effective models for the experimental exploration of the phase-diagram of strongly-interacting matter. We plan to improve our models in various aspects.
First of all we plan to go beyond the present mean-filed approximation, largely employed in the literature, but unable to provide a realistic picture of the hadronic phase, where pions, kaons, protons... are important degrees of freedom contributing to the thermodynamics. Notice that mesons (and also baryons) can be obtained in the PNJL model as solutions of a Bethe-Salpeter equation~\cite{Hansen:2006ee,Costa:2019bua,Torres-Rincon:2015rma}, but one should go one step further and add self-consistently their contribution to the thermodynamic potential~\cite{Wergieluk:2012gd}. Secondly, we plan to add to the models a vector interaction. We should get then a more realistic description of the phase diagram, where presently the value of the nuclear-matter density $n_B\approx 0.16\,{\rm fm}^{-3}$ lies in the phase-coexistence region, representing an unphysical feature of the two models.
Finally, we plan to carry out a deeper study of the first-order transition, investigating aspects like the interface energy between the two phases, the rate of bubble/droplet nucleation and the possible occurrence of spinodal instabilities under the experimental conditions.
We leave the above items for forthcoming publications.

\section*{Acknowledgments}
This work was supported by a research grant by the University of Torino and by an INFN Post Doctoral Fellowship (competition INFN notice n.~18372/2016, RS) as well as by STSM Grants from the COST Action CA15213 ``Theory of hot matter and relativistic heavy-ion collisions'' (THOR). MM and RS thank CFisUC for support and warm hospitality (project UID/FIS/04564/2016 - FCT Portugal). RS thanks Renan C\^amara Pereira, Pedro Costa and Hubert Hansen for valuable discussions and collaboration on related projects.

\bibliography{paper}
%\addbibresource{paper}
\end{document}